\documentclass[11pt]{article}
\usepackage{subcaption}
\usepackage{xr}
\usepackage{changepage}
\usepackage{amsfonts}
\usepackage{geometry}
\usepackage{amsthm}
\usepackage{amsmath}
\usepackage{amssymb}
\usepackage{amsfonts}
\usepackage[authoryear,longnamesfirst]{natbib}
\usepackage{booktabs}
\usepackage{graphicx}
\usepackage{xr}
\usepackage{multirow}
\usepackage{color}
\usepackage{relsize}
\usepackage{titling}
\usepackage[cmyk,x11names]{xcolor}
\usepackage[utf8x]{inputenc}
\usepackage{dsfont}
\usepackage{pgf,pgfplots,import}
\pgfplotsset{compat=newest}
\usepackage[normalem]{ulem}
\definecolor{accentcolor}{RGB}{50, 70, 150}
\usepackage{hyperref}
\hypersetup{pdfborderstyle={},allcolors=accentcolor,colorlinks}

\usepackage{bbm}

\usepackage[
]{ifdraft}

\usepackage[font=normalsize,justification=centering]{caption}
\DeclareCaptionLabelFormat{captionf}{\textsc{#1~#2}}
\usepackage{lscape} 
\usepackage{placeins}
\providecommand{\U}[1]{\protect\rule{.1in}{.1in}}
\usepackage[noamsmath]{kpfonts}
\usepackage{inconsolata}

\usepackage[T1]{fontenc}
\newtheorem{proposition}{Proposition}

\newtheorem{corollary}{Corollary}

{\theoremstyle{definition}
\newtheorem{definition}{Definition}

}

\newtheorem{lemma}{Lemma}

\usepackage[capitalise]{cleveref}
\crefname{observation}{Observation}{Observations}
\crefname{assumption}{Assumption}{Assumptions}
\crefname{proposition}{Proposition}{Propositions}

\definecolor{ColorRed}{RGB}{238,26,28}
\definecolor{ColorBlue}{RGB}{55,126,184}
\definecolor{ColorGreen}{RGB}{77,215,74}
\colorlet{colav}{ColorRed!70!black}
\colorlet{colneu}{ColorBlue!70!black}
\colorlet{colseek}{ColorGreen!70!black}

\begin{document}

\title{\textsc{Who Cares More? Allocation with Diverse Preference Intensities}\thanks{We thank Mariagiovanna Baccara, Kfir Eliaz, Faruk Gul, SangMok Lee, Alessandro Lizzeri, Stephen Morris, and Wolfgang Pesendorfer for very helpful discussions and feedback. Yariv gratefully acknowledges the support of NSF grant SES-1629613.}}
\author{Pietro Ortoleva\thanks{Department of Economics and SPIA, Princeton University.} \quad \quad Evgenii Safonov\thanks{Department of Economics, Princeton University.} \quad \quad Leeat Yariv\thanks{Department of Economics, Princeton University, CEPR, and NBER.}}
\date{\footnotesize{\emph{This version}: \today
}}

\pretitle{\begin{flushleft}\LARGE} 
\posttitle{\end{flushleft}}
\preauthor{\begin{flushleft}\large} 
\postauthor{\end{flushleft}}
\predate{\begin{flushleft}} 
\postdate{\end{flushleft}}
\settowidth{\thanksmarkwidth}{*}
\setlength{\thanksmargin}{-\thanksmarkwidth}

\maketitle

\renewenvironment{abstract}
  {\small
  {\bfseries\noindent{\abstractname}\par\nobreak\smallskip}}
  
\begin{abstract}
\noindent Goods and services---public housing, medical appointments, schools---are often allocated to individuals who rank them similarly but differ in their preference intensities. We characterize optimal allocation rules when individual preferences are known and when they are not. Several insights emerge. First-best allocations may involve assigning some agents ``lotteries'' between high- and low-ranked goods. When preference intensities are private information, second-best allocations always involve such lotteries and, crucially, may coincide with first-best allocations. Furthermore, second-best allocations may entail disposal of services. We discuss a market-based alternative and show how it differs.
\end{abstract}

\thispagestyle{empty}

\vskip 0.2in

\noindent \textbf{Keywords:} Market Design, Mechanism Design, Allocation Problems.

\vskip 0.2in

\noindent \textbf{JEL: }C78, D02, D47.

\newpage \renewcommand{\thefootnote}{\arabic{footnote}} \pagebreak %
\setcounter{page}{1}

\section{Introduction\label{sec:intro}}

In many allocation problems, from public housing to appointment scheduling to some school assignments, individuals largely agree on \textit{ordinal} rankings of goods: prospective public-housing residents usually prefer earlier availability, appointment schedulers tend to prefer sooner service, and parents desire schools in which students perform well on standardized tests. In such settings, heterogeneity appears in \textit{preference intensities}, the \textit{cardinal} rankings. Public housing recipients or appointment schedulers can vary in their urgency, parents may differ in their sensitivity to the ranking of schools their kids attend. Traditional market pricing tools can be used to account for such preference intensities, and lead to efficient allocations. However, in many environments jurisprudence and ethical norms prohibit pricing, making transfers unavailable. How should a social planner allocate services when prices cannot be used? If preferences are not transparent, how can a social planner screen those who care more?

We consider a social planner maximizing utilitarian efficiency and characterize the optimal allocation rules both when preferences are observable and when they are not. When preferences are observable, the unique first-best allocation may probabilistically assign either very high-ranked or very low-ranked services; it may involve a ``lottery.'' When preferences are not observable (private information), the planner faces a screening problem. We show that the unique optimal allocation is fully separating and always involves such lotteries. Crucially, it coincides with the first-best allocation for a non-trivial set of environments. In other cases, first- and second-best allocations differ qualitatively and the second-best allocation may exhibit disposal of services. 

Specifically, we consider an environment in which a continuous supply of goods differing in quality is allocated to a population of agents. Each agent requires only one good. There is a natural ranking of the goods' quality, on which all agents agree. However, agents' preference intensities vary: their valuation of marginal quality changes is different. We capture this disagreement through a natural comparison of marginal utilities, translating into an ordering of utility curvature that can be expressed as an ordering of agents' absolute risk aversion parameters.

This formulation captures many environments. One example is that of homogeneous services supplied over time---identical public-housing units that vary in availability, academic or medical appointments that vary in timing. When agents are exponential discounters, some may be more patient than others, corresponding to well-ranked absolute risk aversion comparisons over timed services. Another example is school choice, where some parents discern small but compelling differences across highly-ranked schools but view lower-ranked schools mostly as an undifferentiated mass; other parents place more equal values on increased school rank. In our baseline setting, agents are of two types: P for prudent or patient, and I for imprudent or impatient.

Suppose first that the social planner observes agents' preferences. This is a convenient technical benchmark that is also relevant for some applications: public housing officials may be informed of home-seekers' circumstances and resulting urgency, academic advisors may be cognizant of students' deadlines when scheduling meetings. We show that the optimal, first-best allocation takes two possible forms. The first has all agents of one type served with goods of the highest quality and all agents of the other type served with goods of lower quality: either all $I$-agents are served with higher-quality goods than $P$-agents, or vice versa. The second possible structure has all $P$-agents, who are more risk averse, served with goods of middling quality, while $I$-agents are served with a ``lottery'' and get, with some probability, either goods of the highest quality or goods of substantially lower quality. 

When are such distributed allocations optimal? Intuitively, the highest-quality goods are more valuable to $I$-agents. When there are many $I$-agents, limited supply of the highest-quality goods implies that some $I$-agents must receive goods that are not of the highest quality. Those $I$-agents who receive lower-quality goods experience a substantially lower utility. For such $I$-agents, a further reduction in quality does not come at a substantial loss. Instead, an equivalent quality reduction for $P$-agents is costlier. It is therefore optimal to serve the highest-quality goods to some of the $I$-agents, then serve the $P$-agents with intermediate-quality goods, and, finally, serve remaining $I$-agents with the lowest-quality goods.

Next, we characterize the optimal allocation when types are unobservable. Extant literature on allocation problems absent transfers commonly assumes complete information of preferences (see our literature review below for a few exceptions). In many cases, however, preference intensities, e.g. urgency of public-housing seekers or appointment schedulers, cannot be observed or confirmed. As is standard in screening problems, the social planner then offers a \textit{menu} of allocations, tailoring the allocations to agent types.

If the first-best allocation serves all agents of one type with goods of uniformly higher quality than those other agents receive, there is no hope for its implementation when types are unobservable: some agent type would have an incentive to mimic the other. However, the first-best
allocation can be incentive compatible if $I$-agents are served probabilistically with either the highest- or the lowest-quality goods. $I$-agents may prefer their allocation to that of $P$-agents as it guarantees them a chance of higher-quality service, which they value greatly. $P$-agents may prefer their assigned allocation since it shields them from the lowest-quality goods. Indeed, we show that, in some cases, the first-best allocation is incentive compatible.

What happens when the first-best allocation is not incentive compatible? We show that the second-best allocation is unique and fully separating. It again takes the form of a distribution over high- and low-quality goods for $I$-agents, and the allocation of goods of contiguous and intermediate quality for $P$-agents. In particular, the pooling allocation, which offers all agents an identical share of the goods' supply and is inherently fair, is never a second-best solution.

Another feature unique to the second-best allocation is that it may exhibit disposal of goods: some agents may receive nothing, even when there is sufficient supply. This occurs when $P$-agents place little marginal value on improving the quality of the goods they receive. In this case, the first-best allocation provides them with the lowest-quality goods. To have an incentive-compatible allocation, the planner needs to eliminate the appeal of the $I$-agents' allocation to the $P$-agents. Certainly, the planner can improve $P$-agents' allocation at the expense of $I$-agents. Alternatively, the planner can worsen $I$-agents' allocation: instead of providing some $I$-agents lower-quality goods, the planner can deny them service altogether. $I$-agents welfare would then be diminished only slightly---their value for the lower-quality goods is relatively low. However, for $P$-agents, such a change can make $I$-agents' allocation substantially less appealing. Denial of service for $I$-agents can therefore be optimal.

Overall, we show that the infinite-dimensional screening problem can be easily reformulated as a simple two-dimensional constrained maximization. Second-best allocations are determined through two levers the planner controls: the set of high-quality goods $I$-agents receive, and the fraction of $I$-agents that are served.

While we describe most of our results for the two-type environment, we show that our qualitative results extend to a setting with an arbitrary number of types. In particular, the first-best solution can be incentive compatible. Furthermore, screening agents for their cardinal preferences is \textit{always} beneficial: the pooling allocation is \textit{never} optimal. 

We also explore a natural alternative to the second-best solution, in the spirit of \cite{HyllandZeckhauser1979}, where agents receive equal shares of the available goods and can trade through a market. The first welfare theorem ensures that induced allocations are Pareto efficient. Nonetheless, we show that resulting allocations may still entail significant efficiency losses relative to second-best allocations.

\paragraph{Related Literature.} A large literature considers screening agents with diverse risk attitudes---through markets, starting from \cite{RothschildStiglitz1976}, or through auctions, starting from \cite{MaskinRiley1984}. While addressing related questions to ours, this literature relies heavily on pricing mechanisms. Therefore, our analysis and the relevant applications are different.

In the context of time preferences, \cite{DellaVignaMalmendier2004} and \cite{EliazSpiegler2006} study screening of time-inconsistent agents. We are not aware of work on screening of time-consistent agents who vary in patience, which our analysis encompasses.

Most work on matching and assignment problems assumes complete information of preferences. \cite{Roth1989} and, more recently, \cite{FernandezYariv2021} illustrate some of the new phenomena that emerge in centralized one-to-one matching markets with incomplete information. Our paper provides insights on the optimal design of allocation protocols in the presence of a particular form of incomplete information.

A recent and growing literature studies dynamic matching and allocations: see \cite{AkbarpourLiGharan2020}, \cite{BaccaraLeeYariv2019}, \cite{BlochCantala2017}, and the survey by \cite{baccara2021dynamic}. \cite{Leshno2017} considers the implications of a desire to speed up assignments on the design of a dynamic allocation procedure. \cite{DimakopoulosHeller2019} show that using wait time as a contractual term can be beneficial in the German market for entry-level lawyers.\footnote{In a somewhat different setting, \cite{ElySzydlowski2020} illustrate how goalposts can be efficiently modified over time in a moral-hazard environment in which tasks take different amounts of time depending on their (uncertain) difficulty. \cite{schummer2020influencing} studies the impacts of risk aversion and impatience on the performance of deferral rights in waiting lists.}

The link between exponential discounting and risk attitudes over lotteries involving timed services has been illustrated by \cite{DejarnetteDillenbergerGottliebOrtoleva2020} and is utilized in some of our examples.

The idea that disposal can be a useful instrument for relaxing incentive constraints is present in other environments, as seen, for example, in \cite{AlatasETAL2016} in the context of applications for aid programs and \cite{AustenSmithBanks2000} in the context of cheap talk. Technically, our observation regarding the optimal use of probabilistic allocations relates to \cite{GauthierLaroque2014}, who provide general necessary and sufficient conditions for stochastic optimization solutions.\footnote{When a buyer and seller have correlated valuations of a good, \cite{kattwinkel2020allocation} shows that the seller's optimal mechanism may involve randomization. Reminiscent of some of our results, with positive correlation, the good may not be allocated to a higher-value buyer with higher probability.}

Finally, we discuss a market-based implementation that is inspired by \cite{HyllandZeckhauser1979} and work that followed.

\section{The Allocation Problem}
\label{sect:Model}

\subsection{Setup}

We study the allocation of a continuum of goods to agents of heterogeneous preferences.

\paragraph{Goods.} Goods are characterized by a one-dimensional attribute $x \in [0,X]$. They can stand for public housing units available at different times, doctor appointments that vary in physician's expertise or date of service, schools that vary in quality, \textit{etc}. The available supply of different goods is captured by a strictly positive continuous density $f$ over $[0,X]$, with cumulative distribution $F$. 

\paragraph{Agents.} We start by considering two types of agents. We extend our analysis to $N$ types in Section \ref{sect:Ntypes}. Agents are of type $P$ and $I$, with strictly positive masses $\mu _{P}$ and $\mu_{I}$, respectively. Each agent demands one unit of the good. For presentation ease, we assume there is sufficient supply; that is, $F(X) \geq \mu_{P}+\mu_{I}$.\footnote{When supply in insufficient, the analysis is broadly similar but requires consideration of multiple cases depending on the severity of the goods' scarcity.} 

$P$- and $I$-agents have utilities $u_{P}$ and $u_{I}$ over $\mathbb{R}_{+}\cup{\diamond}$, respectively, where $\diamond$ denotes receiving none of the goods. Agents agree on the ordinal ranking of goods: both $u_{P}$ and $u_{I}$ are strictly decreasing and twice continuously differentiable over $\mathbb{R}_{+}$.\footnote{Goods' labels can therefore be thought of as a continuous rank, with $x=0$ representing the most-preferred good, and $x=X$ representing the least-preferred good.} They also rank any of the goods strictly higher than $\diamond$. Without loss of generality, we posit $u_{P}(\diamond)=u_{I}(\diamond)=0$. We also assume that as goods' quality deteriorates to $\infty$, their value approaches that of not receiving any good: $\lim_{x\to\infty}u_{P}(x)=\lim_{x\to\infty}u_{I}(x)=0$.

While agents have the same \textit{ordinal} ranking of goods, they disagree on \textit{cardinal} assessments. Specifically, we assume that for all $x \in \mathbb{R}_+$,
\begin{equation}
\quad \quad  \frac{u_{P}^{\prime \prime }(x)}{u_{P}^{\prime }(x)}>\frac{u_{I}^{\prime
\prime }(x)}{u_{I}^{\prime }(x)}. 
\label{eq:Utility_Ranking}
\end{equation}%
That is, since agents' utilities are decreasing, $P$-agents have greater absolute risk aversion than $I$-agents. Put another way, $I$-agents have ``more convex'' or ``less concave'' utility functions. We emphasize that this comparison is \textit{relative}---both utilities can be convex, concave, or change curvature within the domain. Below we discuss a few examples that fit our framework. 

\paragraph{Allocations.} A \textit{lottery} is a probability distribution on $[0,X]\cup \{\diamond \}$. Since the supply $f$ has continuous density, we focus on lotteries that have no mass points on $[0,X]$. A lottery $q$ then specifies a density over $[0,X]$. Abusing notation, we denote by $q(\diamond)$ the remaining probability: $q(\diamond):=1-\int_{[0, X]} q(x)\mathrm{d}x$. For any measurable $A \subseteq [0,X]\cup \{\diamond \}$, we denote $q(A):=\int_{A\backslash \{\diamond\}}q(x)\mathrm{d}x + \mathbbm{1}_{\diamond \in A} q(\diamond)$. 

An \textit{allocation} is a pair $(q_P, q_I)$, where $q_P$ and $q_I$ are  $P$-agents' and $I$-agents' allocation, or lottery, respectively. An allocation is \textit{feasible} if goods assigned are available: for (almost) all $x \in [0, X]$
\begin{equation*}
\mu_P q_P(x)+ \mu_I q_I(x) \leq f(x). 
\end{equation*}

While feasibility is a natural requirement, in some applications the planner may be able to weaken it by lowering the quality of some good: for example, for allocations of goods over time, the planner may be able to store some of the unassigned goods, increasing availability in future periods. This naturally expands the set of feasible allocations. As it turns, allowing for such expansion does not alter our results. We discuss details in our Conclusions and Online Appendix.

Finally, the following notation will be useful: for any measurable $A, B \subseteq [0, X] \cup \{\diamond\}$, $A \triangleleft B$ denotes the case in which any element in $A$ has strictly higher quality than any element of $B$; that is, $x<x^{\prime}$ for any $x\in A$, $x^{\prime}\in B \backslash \diamond$, and $\diamond \notin A$. 

\paragraph{Expected Payoffs.} Agents evaluate their allocation using expected utility. That is, for $k\in \{P,I\}$, $k$-agents' utility from allocation $q$ is given by:\footnote{Recall that $u(\diamond)$ has been normalized to 0, and therefore does not appear in the expression.}
$$ 
V_{k}(q)=\overset{X}{\underset{0}{\int }}u_k(x)q_k(x)\mathrm{d}x.
$$

\subsection{Examples}

\paragraph{Assignment Problems and Risk Attitudes.}

The set $[0, X]$ may stand for different qualities of goods assigned at the same time: the quality of schools in a school-choice problem, of houses in a real-estate market, of doctors in healthcare allocation problems, and so forth. In many applications, ordinal preferences of agents are highly correlated---schools may have publicly available rankings and houses may have features desired by most. Our model focuses on the case in which ordinal preferences are identical, but cardinal preferences are different. Some individuals have large marginal returns for improved quality, others less so. In this context, our condition (\ref{eq:Utility_Ranking}) on utilities fits many common functional forms: for example, CRRA or CARA utilities of varying parameters are ranked via our condition. 

Our condition on utilities can also be read directly in terms of heterogeneity in risk attitudes. 
The planner's problem can thus be seen as one of screening different risk attitudes over quality. As discussed above, most extant work considers screening of agents over risk attitudes using pricing mechanisms, which may not be germane to many of the applications we consider.

\paragraph{Timing of Goods and Services.}\label{sect:Timing} A natural application is to homogeneous goods available at different dates, where $x\in [0,X]$ denotes a delivery time. Agents have instantaneous utility for the good normalized to $1$, but discount at different rates: $r_{P}$ for the more patient $P$-agents and $r_{I}$ for the more impatient $I$-agents, where $0<r_{P}<r_{I}$.

Many examples fit this application: scheduling appointments or services, public housing available over time, \textit{etc}. The planner's problem is then one of general scheduling, determining how a given supply of timed appointments, houses, and the like should be allocated when agents have different discount rates, or \textquotedblleft urgency\textquotedblright.

\cite{DejarnetteDillenbergerGottliebOrtoleva2020} illustrate that exponential discounters are strictly risk seeking over the date at which they receive the good. Indeed, impatient $I$-agents are strictly more risk seeking than patient $P$-agents: since they care more about immediate service, they are more willing to take lotteries with larger spread that provide either highly-desirable quick service, or greatly delayed service. Formally, we have $e^{-r_{P}x}=\left(e^{-r_{I}x}\right) ^{\frac{r_{P}}{r_{I}}}$, and utilities satisfy our assumption (\ref{eq:Utility_Ranking}).

While exponential discounting is a natural case, our analysis applies also to other forms of utility loss from waiting, as long as (\ref{eq:Utility_Ranking}) holds. For example, agents could be present-biased, discounting can be hyperbolic or quasi-hyperbolic, and so on.

\section{First-best Allocations}\label{sect:FirstBest}

We begin with the case in which types are observable. This is not only a natural theoretical benchmark, it also speaks to various applications: healthcare systems may be able to assess patients' urgency, social services may be able to gauge individuals' immediate needs for public housing. 

The planner's problem is to find a feasible allocation $(q_P, q_I)$ that maximizes the weighted utilitarian welfare function $W$:
\begin{equation}
W(q_P, q_I):= \alpha \mu_P V_P(q_P) + (1- \alpha) \mu_I V_I(q_I),
\end{equation}
where $\alpha\in (0,1)$ denotes the weight placed on $P$-agents' expected utility. While we consider a general model, $\alpha=\frac{1}{2}$ is a natural special case in which agents are valued equally. We call a solution to this problem the \textit{solution to the planner's problem} or the \textit{first-best}. 

Denote by $\overline{X}:= F^{-1}(\mu_I+\mu_P)$ the lowest quality needed to exhaust demand if only the best qualities are used. Since we assume sufficient supply, in the first-best allocation all agents receive a good and only highest-quality goods are used. Thus, the trade-off  that the planner needs to resolve pertains only to the allocation of goods within $[0,\overline{X}]$. Which goods should go to $I$-agents and which to $P$-agents? The trade-off is captured by the difference between agents' utilities and summarized by the function $g: \mathbb{R}_+ \cup \{\diamond\} \to \mathbb{R}$ defined as
\begin{equation}
g(x) := \alpha u_P(x) - (1-\alpha) u_I(x). 
\end{equation}
The planner would like to assign a good of quality $x$ to $P$-agents when $g(x)$ is high, and to $I$-agents when $g(x)$ is low. The characterization of the first-best solution thus tracks the shape of $g$. 

\begin{lemma}\label{lemma:FunctionG}
The function $g$ is single-peaked and strictly quasi-concave. 
\end{lemma}

For intuition, take our example of homogeneous goods over time, where utility is of the form $e^{-r_jx}$. If equal weights are placed on both types, then $g(0)=0$---there is no valuation difference between types at time zero. The difference $g(x)$ becomes arbitrarily small for very late delivery time. However, $g$ remains strictly positive in between and there is a unique maximum of $g(x)$ for some strictly positive $x\in [0,X]$. 

The first-best solution can be characterized using the function $g$. Recall that $\overline{X}$ is the lowest quality of goods assigned. For any $j=I,P$, denote by $\overline{X}_j:=F^{-1}(\mu_j)$ the lowest quality of goods assigned to $j$-agents if $j$-agents were to receive the best-quality goods. Naturally, $\overline{X}_j < \overline{X}$ for $j=I,P$. 

\begin{figure}[t!]
    \centering
    \hspace*{-2cm} \includegraphics[width=18cm]{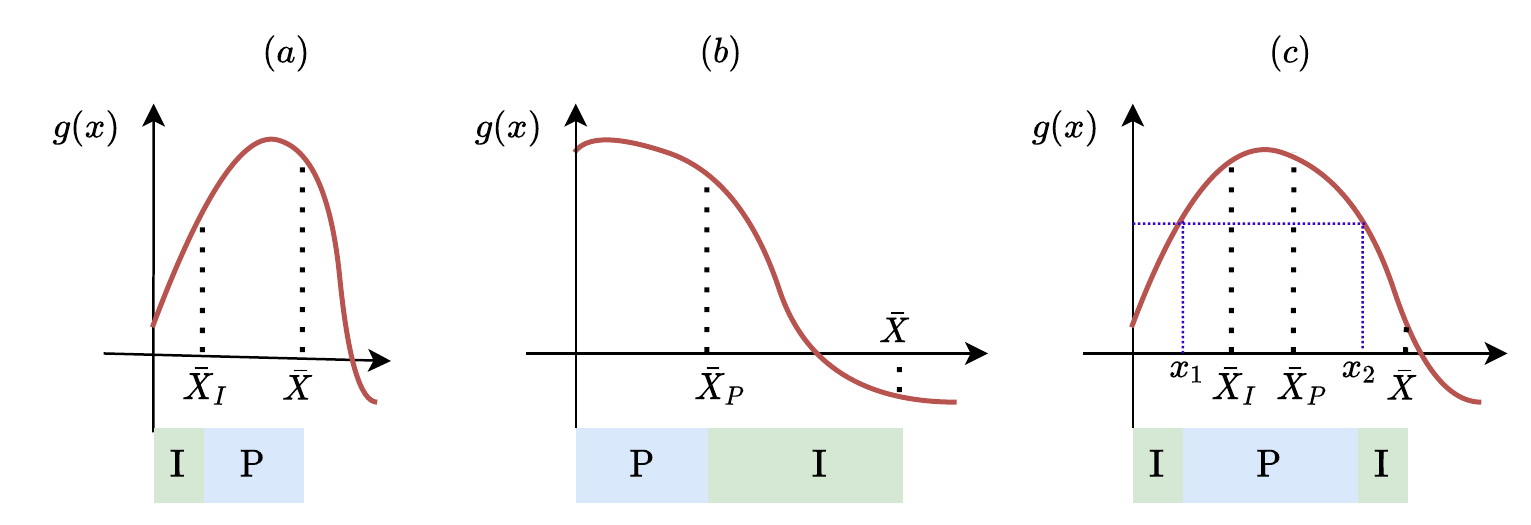}
    \caption{Computing the first-best allocation.}
		\label{fig:gFunction}
\end{figure}

Now consider $g(\overline{X}_{I}), g(\overline{X}_{P})$, and $g(\overline{X})$. Suppose first that $g(\overline{X}_{I})< g(\overline{X})$. This is illustrated in panel (a) of Figure \ref{fig:gFunction}. The planner assigns goods to $I$-agents when $g$ is low and to $P$-agents when $g$ is high. This is achieved by exhausting $I$-agents' demand with the highest-quality goods, namely allocating them goods in $[0, \overline{X}_{I}]$. Lower-quality goods in $[\overline{X}_{I}, \overline{X}]$ are then allocated to $P$-agents. We term the resulting allocation structure IP. 

Suppose $g(\overline{X}_{P}) \leq g(0)$, as depicted in panel (b) of Figure \ref{fig:gFunction}. Because $g$ is single peaked, $g$ is decreasing to the right of $\overline{X}_{P}$. Thus, the highest values of $g$ are achieved to the left of $\bar{X}_p$, and it is optimal to exhaust $P$-agents' demand with the highest-quality goods, those with quality in $[0,\overline{X}_{P}]$. Lower-quality goods, in $[\overline{X}_P,\overline{X}]$ are assigned $I$-agents. We term this allocation structure PI.

In the remaining case, $g(\overline{X}_I)>g(\overline{X})$ and $g(\overline{X}_{P})>g(0)$, as in panel (c) of Figure \ref{fig:gFunction}. The highest levels of the $g$ occur between $\overline{X}_I$ and $\overline{X}$. The optimal allocation then has $P$-agents served with goods in $(x_{1},x_{2})\subsetneq (0,\overline{X})$, with $g(x_{1})=g(x_{2})$ and $F(x_{2})-F(x_{1})=\mu_{P}$.\footnote{If $g(x_{1})<g(x_{2})$, since $g(x)$ is continuous, the planner would benefit from serving a small mass of $P$-agents with goods of qualities just below $x_{2}$ instead of qualities just below $x_{1}$. A similar argument follows if $g(x_{1})>g(x_{2})$.} The interval $(x_{1},x_{2})$ is uniquely determined by these two constraints. The resulting allocation has $I$-agents served with relatively high- and low-quality goods, while $P$-agents are served with intermediate-quality goods. We term this allocation structure IPI.

The following proposition summarizes our discussion. For any $A \subset [0,X]$, denote by $f ~\big\vert~ A $ the allocation that assigns the full supply available in $A$.\footnote{Formally, $\left( f ~\big\vert~ A \right)~ (x) = f(x) \cdot \mathbbm{1}\{x \in A \} \big/ \int_{y \in A}f(y)dy$.} 

\begin{proposition}[First-Best]
There exists a unique first-best allocation $(q_{P},q_{I})$. Moreover:

\begin{enumerate}

\item If $g(\overline{X}_{I}) \leq g(\overline{X})$, then $I$-agents are assigned higher-quality goods than $P$-agents:
\begin{equation}
q_{P}~=~f ~\big\vert~ [\overline{X}_I, \overline{X}] \quad \quad q_{I}~=~f ~\big\vert~ [0, \overline{X}_I]  \quad \quad \quad \quad (IP~~structure)
\notag 
\end{equation}

\item If $g(\overline{X}_{P}) \leq g(0)$, then $P$-agents are assigned higher-quality goods than $I$-agents:
\begin{equation}
q_{P}~=~f ~\big\vert~ [0, \overline{X}_{P}] \quad \quad q_{I}~=~f ~\big\vert~ [\overline{X}_{P}, \overline{X}]  \quad \quad \quad \quad (PI~~structure)
\notag
\end{equation}

\item Otherwise, $P$-agents are assigned goods with quality in between that of those assigned to $I$-agents:
\begin{equation}
q_{P}~=f ~\big\vert~[x_{1},
x_{2}], \quad \quad q_{I}~=f ~\big\vert~[0,x_{1}]\cup
[x_{2},\overline{X}]  \quad \quad (IPI~~structure)
\notag
\end{equation}%
where $0 <x_{1} < x_{2} < \overline{X}$ is the unique solution of $g(x_{1})=g(x_{2})$ and $F(x_{2})-F(x_{1})=\mu_{P}$.

\end{enumerate}
\end{proposition}

Our discussion also suggests how, in general, neither of the three cases---IP, PI, or IPI---is knife-edge. This is immediate to see in our example of homogenous goods over time. When $\alpha\leq \frac{1}{2}$, we have $g(x)>g(0)=0$ for all $x>0$, and the PI structure is never optimal. However, there are positive measures of discount factors, welfare weights, and masses of $I$- and $P$-agents, for which the IP structure or the IPI structures are optimal. Similarly, if $\alpha> \frac{1}{2}$, then for any $r_{P}<r_{I}$, we have $g(x)<g(0)$ for large enough $x$. Thus, the PI structure is optimal when a large enough mass of $P$-agents is present.\footnote{Arguments of robustness can be made stronger. We could endow both the set of viable utilities and the set of supply functions with a specific metric on each dimension of the problem (e.g., the sup-norm for utilities or supply functions) and show that, for any of the three plausible structures, there exists an open set of parameters---masses, utilities, and welfare weights for either agent type, as well as supply functions---so that the specific structure is optimal within that set.}

\paragraph{Comparative Statics.}

How does the solution structure change with the underlying parameters of the environment? Fix the overall mass of agents and assume first that the maximizer of $g$, call it $x^*$, is in $(0,\overline{X})$. 

If $g(0)< g(\overline{X})$, then $g(x)>g(0)$ for all $x \in (0,\overline{X}]$ and only cases (1) and (3) of Proposition 1 can occur. When $\mu_I$ is small, $\overline{X}_I$ is close to $0$ and an IP structure is optimal. As $\mu_I$ increases, $\overline{X}_I$ increases and, for some level of $\mu_I$, $g(\overline{X}_I)=g(\overline{X})$. Any further increase of $\mu_I$ yields an IPI structure as the solution. 

Similarly, if $g(0)> g(\overline{X})$, then $g(x)>g(\overline{X})$ for all $x \in [0, \overline{X})$ and only cases (2) and (3) of Proposition 1 can occur. When $\mu_I$ is small, $\overline{X}_P$ is close to $\overline{X}$ and a PI structure is optimal. As $\mu_I$ increases, or $\mu_P$ decreases, $\overline{X}_P$ decreases. As before, for some level of $\mu_I$, $g(\overline{X}_P)=g(0)$. Any further increase of the mass of $I$-agents yields an IPI structure as optimal.

If $x^*$ is either $0$ or $\overline{X}$, then $g(x)$ is monotone over $[0, \overline{X}]$, and changes in the relative masses of agent types do not affect the structure of the first-best solution: if $x^*=0$, we have a PI structure; if $x^*=\overline{X}$, an IP one.

\begin{corollary}
Fix the overall mass of agents $\mu_I+ \mu_P$. Let $x^{*} := \underset{x \in [0,\overline{X}]}{\textrm{arg~max}} ~g(x)$, then:
\begin{enumerate}
    \item If $x^*\in (0,\overline{X})$ and $g(0)< g(\overline{X})$, then there exists $\Tilde{\mu}_I$ such that, for all $\mu_I < \Tilde{\mu}_I$, the first-best solution exhibits the IP structure and, for all $\mu_I>\Tilde{\mu}_I$, the first-best solution exhibits the IPI structure.
    \item If $x^*\in (0,\overline{X})$ and $g(0)> g(\overline{X})$, then there exists $\overline{\mu}_I$ such that, for all $\mu_I < \overline{\mu}_I$, the first-best solution exhibits the PI structure and, for all $\mu_I>\overline{\mu}_I$, the first-best solution exhibits the IPI structure.
    \item If $x^*\in \{0,\overline{X}\}$, the first-best solution exhibits the same structure, either PI or IP, regardless of the relative masses of agents.
\end{enumerate}
\end{corollary}

\begin{figure}
	\centering
	\includegraphics[width=7cm]{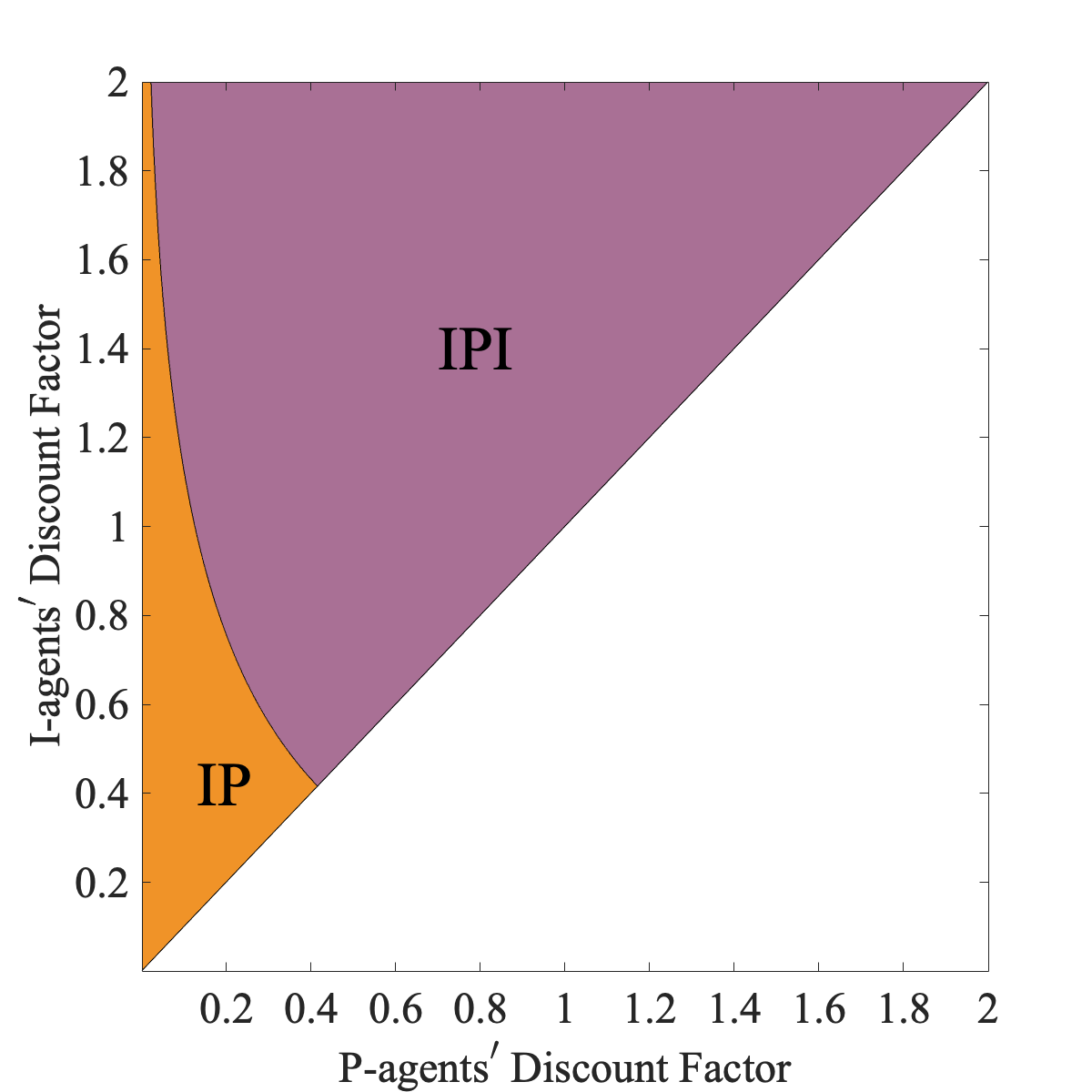}
	\caption{First-best allocations in the timed homogeneous-good example as a function of agents' discounts.}
		\label{fig:CompStatFirstBest}
\end{figure}

We can also analyze the impact of the relative curvature of utilities. Considering again our example of homogeneous goods over time, we have $g(x)=\alpha e^{-r_{P}x}-(1-\alpha)e^{-r_{I}x}$. If agents are weighted equally ($\alpha=\frac{1}{2}$), then $g(0)<g(x)$ and $g$ is maximized at $x^*=\frac{\ln{r_I}-\ln{r_P}}{r_I-r_P}$. When $r_P$ is low, so that $P$-agents are patient and their utilities are fairly flat over delivery times, serving the impatient $I$-agents first is optimal and the first-best solution exhibits an IP structure. As $r_P$ grows, the maximizing $x^*$ becomes small, and the first-best solution exhibits an IPI structure. Intuitively, $I$-agents are impatient enough so that, beyond a certain early time, further delays do not entail significant welfare costs. However, for the more patient $P$-agents, minor delays are not as costly, but substantial delays are. Figure \ref{fig:CompStatFirstBest} depicts the regions of discount factors corresponding to each first-best structure, IP or IPI, assuming equal mass of agents and uniform supply.\footnote{In particular, $\alpha=\mu_I=\mu_P=\frac{1}{2}$, and supply with total mass $3/2$ is distributed uniformly over $[0,5]$.}

\section{Incentive-Compatible Mechanisms}\label{sect:SecondBest}

We now turn to the case in which types are not observable: in some cases, urgency for or preferences over public-housing units are difficult to ascertain; for mundane treatments, medical offices may be unable to assess the necessity for quick attention; universities may be unable to assess preference intensities over dorm rooms; school systems may not be privy to parents' or students' strength of preferences for one school over the other; \textit{etc}. As before, the welfare-maximizing mechanism designer would like to associate a different lottery to each agent type. However, now the choice of who receives which lottery is effectively done by the agents---they report their type, which yields an allocation. The designer then needs to take care of additional constraints, corresponding to agents picking lotteries tailored for them. This boils down to a standard screening problem. We call it the mechanism-designer problem: 
\begin{equation}
\begin{array}{lrlll}
	& \multicolumn{4}{c}{ \underset{(q_P, q_I)}{\text{max}} \ \ \alpha \mu
_{P} V_{P}(q_P)+(1-\alpha )\mu _{I}V_{I}(q_I) \text{ \ \ \ such that }}  \label{eq:Mech_Des_Problem}\\
(IC_{kj}) & \ \ \ \ \ \ \ \ \ V_{k}(q_k) &  \geq  \ V_{k}(q_j) & \ \ \ \ \ \ \forall k, j \in \{P, I\} \\ 
(\text{Feasibility}) & \ \ \ \ \ \ \ \ \ \mu_P q_P(x)+ \mu_I q_I(x) & \leq \ f(x) & \ \ \ \ \ \ \forall x \in [0, X].
\end{array}
  \notag
\end{equation}

\indent Like the planner, the mechanism designer chooses a menu $(q_P, q_I)$, where $q_k$ is the lottery designed for $k$-types, $k=P,I$. However, the mechanism designer needs to respect additional incentive-compatibility constraints, $IC_{kj}$, with $k,j \in \{P,I\}$, ensuring that $k$-agents do not want to emulate $j$-agents.

Standard arguments guarantee that a solution to the mechanism designer's problem always exists (see Appendix). We refer to this solution as the \textit{second-best}. 

\subsection{Can the First-Best be Achieved?} 

We begin our analysis by asking: Can first-best allocations be achieved when types are unobservable? Solutions with an IP or a PI structure are naturally not incentive compatible: agents receiving lower-quality goods would benefit from misreporting their type---no matter the curvature of the utility function, their allocation is first-order stochastically dominated.

When the first-best is of the IPI form, however, allocations are no longer ranked in terms of first-order stochastic dominance; which one is preferred depends on agents' utility functions. We assumed that $u_I$ decreases more sharply at higher qualities than $u_P$. Thus, $I$-agents may be willing to accept the risk of getting lower-quality goods in exchange for the chance of getting higher-quality ones; $P$-agents, instead, may prefer a more ``balanced'' allocation. Put differently, in an IPI allocation, the lottery for $I$-agents is more risky than the one for $P$-agents. Because $I$-agents are more risk seeking, some IPI allocations can be incentive compatible.

\begin{proposition}
	There exists a non-degenerate closed interval of positive weights $\alpha$ for which the first-best allocation is incentive compatible.
\end{proposition}

The proposition shows that there are non-trivial cases in which the first-best is achievable via an incentive compatible mechanism.\footnote{In fact, the proof shows that there exists a closed interval within $(0,1)$ such that the first-best allocation is incentive compatible \textit{if and only if} the welfare weight $\alpha$ is in that interval.} For the time-discounting case, for example, for any supply function, there is an open set of the parameters---planner's weights, discount factors, masses of types---for which the first-best allocation coincides with the second-best allocation.

\subsection{Features of Incentive-Compatible Mechanisms}

To characterize the solution of the mechanism designer's problem, we identify several necessary features it must exhibit.

\paragraph{No inverted spread and its implications.} We already saw that allocations of the form IPI may be incentive compatible since they exhibit a larger ``spread'' in the allocation tailored to the more risk-seeking $I$-agents. We now define the mirror image, what we call an``inverted spread,'' where $I$-agents receive goods of quality in-between that provided to $P$-agents.

\begin{definition}
	An allocation $(q_P, q_I)$ exhibits an \emph{inverted spread} if there exist $A,B,C\subseteq \lbrack 0,X)\cup \{\diamond \}$ such that $A \triangleleft B \triangleleft C$ and $q_{P}(A),q_{I}(B),q_{P}(C)>0$.
\end{definition}

Our first step illustrates inverted spreads never occur in the second-best solution. 

\begin{lemma}\label{lemma:NoWrongSpreads}
	Solutions of the mechanism designer's problem never exhibit an inverted spread. 
\end{lemma}

\begin{figure}
	\centering
	\includegraphics[height=4.5cm]{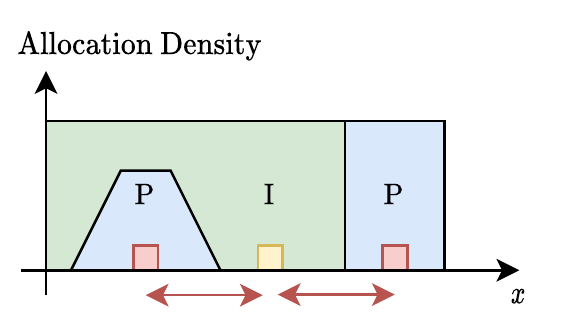}
	\caption{Intuition for Lemma \ref{lemma:NoWrongSpreads}, `No inverted spread.'}
		\label{fig:NoWrongSpread}
\end{figure}

The intuition is illustrated in Figure \ref{fig:NoWrongSpread}. Suppose an allocation exhibits an inverted spread. Fix one lottery corresponding to intermediate-quality goods within the support of $I$-agents' allocation---the small yellow rectangle. There exist a lottery that provides, with some probability, each of two qualities within the support of $P$-agents' allocation---the small red rectangles in the figure---such that $P$ agents are indifferent between the two lotteries. Since $I$-agents are more risk loving, they \textit{strictly} prefer the lottery that has more extreme qualities in its support. Thus, the original allocation cannot be optimal: the designer can ``swap'' some of the goods, assigning a small amount of intermediate-quality goods to $P$-agents, and more spread-out quality goods to $I$-agents. This increases welfare preserving incentive compatibility. 

Three implications follow. First, all solutions must be ``fully'' separating: a given quality level is never assigned to both types (except for measure-zero sets). If it did, the allocation would exhibit an inverted spread. This immediately rules out the pooling allocation---placing uniform probability over qualities from 0 to $\overline{X}$---as a possible solution.\footnote{Formally, the pooling allocation $\big(q^{\mathrm{pool}}_P, q^{\mathrm{pool}}_I\big)$ is defined by  $q_{I}^{\mathrm{pool}}=q_{P}^{\mathrm{pool}}=f ~\vert~ [0,\overline{X}]$.}

A second implication is that, in any solution, the two IC constraints cannot bind at the same time. Indeed, suppose both $IC$ constraints bind. Then, both types of agents must be indifferent between the allocations. As Expected Utility is linear in probabilities, agents must also be indifferent between any convex combination of the allocations. Thus, any such convex combination, and in particular the pooling allocation, is also a solution, in contradiction.

The third implication is that all $P$-agents receive a good: $q_P(\diamond)=0$. Certainly, if all $P$-agents receive lower-quality goods than $I$-agents or no goods at all, incentive compatibility is violated. Suppose some $I$-agents receive lower-quality goods than some $P$-agents and that other $P$-agents are denied service. That would yield an inverted spread, which cannot occur.

\begin{corollary}\label{cor:NoBothIC}
	If $(q_P, q_I)$ is a solution of the mechanism designer's problem, then $\mathrm{supp}(q_P) \cap$ $\mathrm{supp}(q_I)$ $\cap$ $[0,X]$ has measure zero. Moreover, $IC_{IP}$ and $IC_{PI}$ are not both binding.
\end{corollary}

\paragraph{$P$-agents served in ``one-block.''} The next step illustrates that $P$-agents must be served in one continuous block. That is, there exists an interval of qualities such that all supply in that interval is given to $P$-agents, and this exhausts their demand.  

\begin{lemma}\label{lemma:PtypesInOneBlock}
If $(q_P, q_I)$ is a solution of the mechanism designer's problem, then there exists $x_1, x_2 \in [0, X]$ such that $F(x_{2})-F(x_{1}) = \mu_{P}$, and $q_{P} = f ~\vert~ [x_{1},x_{2}]$.
\end{lemma}

Intuitively, $P$-agents are more risk averse, and are therefore optimally served contiguously. Because there is no inverted spread, no $I$-agents are served in between. The lemma shows that there are no ``gaps'' with goods left unassigned. To see why, suppose that a solution with such a gap exists and recall that IC-constraints cannot bind simultaneously. 

\begin{figure}
	\centering
 \hspace*{-1cm} \includegraphics[height=4.5cm]{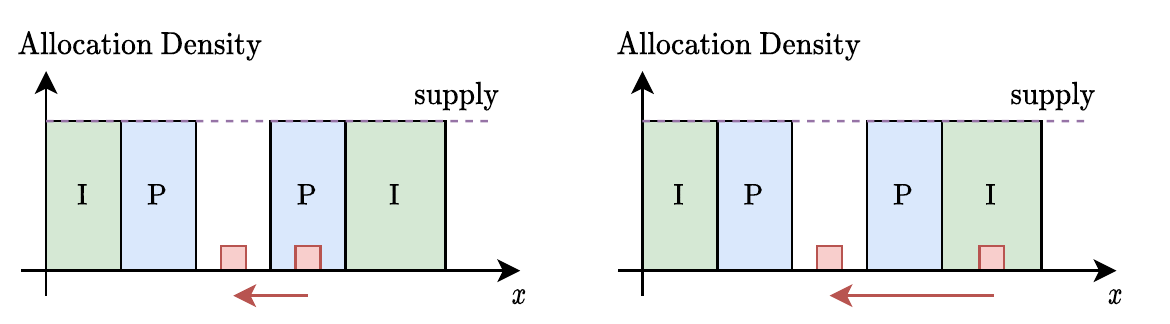}
	\caption{Intuition of Lemma \ref{lemma:PtypesInOneBlock}, $P$-agents served in one block.}
		\label{fig:IntuitionLemmaPtypes}
\end{figure}

If $IC_{IP}$ does not bind, consider a small quantity of the lower-quality goods served to the $P$-agents and swap it for a small quantity of equal mass of the higher, unassigned quality goods in the ``gap''---see the left panel of Figure \ref{fig:IntuitionLemmaPtypes}. This would improve $P$-agents' allocation and welfare. Since $IC_{IP}$ does not bind, for a small enough mass of goods swapped, incentive compatibility for $I$-agents is preserved, in contradiction.

Suppose $IC_{IP}$ binds, so that $IC_{PI}$ does not. Consider a small quantity of lower-quality goods served to $I$-agents and swap it for an equal quantity of higher, unassigned quality goods in the ``gap''---see the right panel of Figure \ref{fig:IntuitionLemmaPtypes}. Such a swap improves $I$-agents' utility and, if it entails a small enough mass, does not violate incentive compatibility for $P$-agents, again in contradiction. 

\paragraph{Disposal of goods for $I$-agents.} We established that all $P$-agents are served in one continuous block. We now discuss the use of disposal, when some agents---namely, the $I$-agents---are denied service despite the availability of goods.

\begin{definition}
An allocation $(q_{P},q_{I})$ exhibits \emph{disposal} for agent of type $k\in \{P,I\}$ if there exist sets $A,B\subset \lbrack 0,X]\cup \{\diamond \}$  such that $A \triangleleft B$, 
$f(A)-\mu _{P}q_{P}(A)-\mu _{I}q_{I}(A)>0$, and $q_{k}(B)>0$.
\end{definition}

We show that the only possible form of disposal is one that precludes some $I$-agents from receiving goods.

\begin{lemma}\label{lemma:DelayToInfinityForI}
If $(q_P, q_I)$ is a solution of the mechanism designer's problem, then:
\begin{enumerate}
    \item $(q_P, q_I)$ does not exhibit disposal for $P$-agents;
    \item If $(q_P, q_I)$ exhibits disposal for $I$-agents, then $q_I(\diamond)>0$. Moreover, there are no $A, B \subset [0, X]$ such that $A \triangleleft B$,  $\mu_I q_I(A)+\mu_P q_P(A) < f(A)$, and $q_I(B)>0$.
\end{enumerate}
\end{lemma}

\begin{figure}
	\centering
	\includegraphics[height=4.5cm]{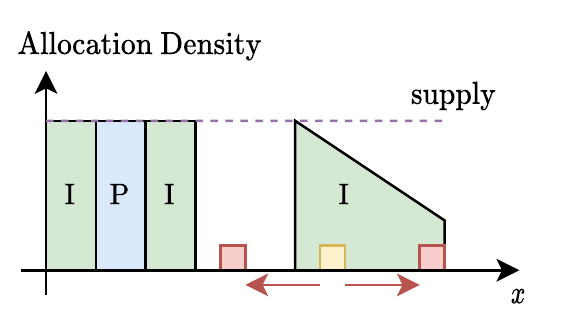}
	\caption{Intuition for Lemma \ref{lemma:DelayToInfinityForI}, only disposal can be denial of goods for $I$-agents.}
		\label{fig:IntuitionDelayInfinity}
\end{figure}

Suppose a second-best solution exhibits disposal such that some $I$-agents are served with a lower-quality good when higher qualities are available, as depicted in Figure \ref{fig:IntuitionDelayInfinity}. We can use similar arguments to those already used to show there are no inverted spreads. Consider two lotteries. One lottery is supported by unassigned higher-quality goods and available lower-quality goods---the small red rectangles in the figure. A second lottery is supported by intermediate-quality goods within the support of $I$-agents---the yellow rectangle in the figure. We can set the probabilities so that $P$-agents are indifferent. Since $I$-agents are more risk loving, they strictly prefer the first, more diverse lottery. As before, a swap of these lotteries increases welfare while preserving incentive compatibility, contradicting optimality.

This line of arguments implies that if the mechanism designer utilizes disposal, she would do so even if an additional supply of goods of quality lower than $X$ were added. We show below that disposal may indeed occur in the second-best allocation.

\subsection{The Optimal Mechanism}

We are now ready to state the characterization of the second-best allocation.

\begin{proposition}\label{th:CharacterizationSecondBest}
	 There exists a unique solution of the mechanism designer's problem, given by $$
	 q_P~=~f ~\big\vert~ [x_1,x_2]
	 $$
	$$
    q_I~=~(1-\beta) \cdot f ~\big\vert~ [0,x_1] \cup [x_2,x_3]~+~\beta \cdot \delta_{\diamond}
	$$
	where $\beta \in [0,1)$, $0 < x_{1} < x_{2} \leq x_{3} \leq X$, $F(x_2)-F(x_1)=\mu_P$, and $F(x_1)+ \left(F(x_3)-F(x_2)\right)=(1-\beta) \mu_I$.
\end{proposition}

The proposition shows that the second-best allocation is unique, and always takes the IPI structure, with the caveat that some $I$-agents may not be served at all. $P$-agents' demands are exhausted over an interval $[x_1, x_2]$, where $x_1>0$. $I$-agents are served with high-quality goods in $[0, x_1)$ and lower-quality goods in $[x_2, x_3]$. These may not exhaust their demands and with probability $\beta$ they receive no good. 

The proposition also illustrates that the original, infinite-dimensional problem is effectively reduced to a two-dimensional problem. There are only two levers the mechanism designer can use. The first is $x_1$, which specifies the quantity of the highest-quality goods $[0,x_1]$ distributed to $I$-agents. Once $x_1$ is set, the length of the interval $[x_1, x_2]$ follows: the supply of goods over that interval must coincide with the $P$-agents' mass. Any choice of $x_1$ therefore uniquely pins down $x_2$. The second lever is then $\beta$, or equivalently $x_3$, which governs the probability that $I$-agents are served with a good and, in turn, the interval of lower-quality goods $[x_2, x_3]$ they are served with. 

Why might it be optimal to dispose of some goods and not serve $I$-agents? Consider a candidate allocation in which all $I$-agents are served and receive goods in $[0, x_1]$ and $[x_2, x_3]$, while $P$-agents receive goods in $[x_1, x_2]$. Suppose it violates $IC_{PI}$: $P$-agents prefer $I$-agents' allocation. There are two adjustments the mechanism designer can contemplate. She can make $P$-agents' allocation more attractive, reducing $x_1$ and $x_2$ to improve the quality of goods they receive. Alternatively, the designer can reduce the desirability of $I$-agents' allocation by decreasing the quality of goods they receive, potentially precluding some $I$-agents from goods altogether. Taking away goods of very low quality from $I$-agents has little impact on their welfare. But it may substantially reduce the appeal of the $I$-agents' allocation for $P$-agents. Due to the asymmetry in how agents evaluate lower-quality goods, it may be efficient for the mechanism designer to use disposal. 

\begin{figure}
	\centering
	\includegraphics[width=7cm]{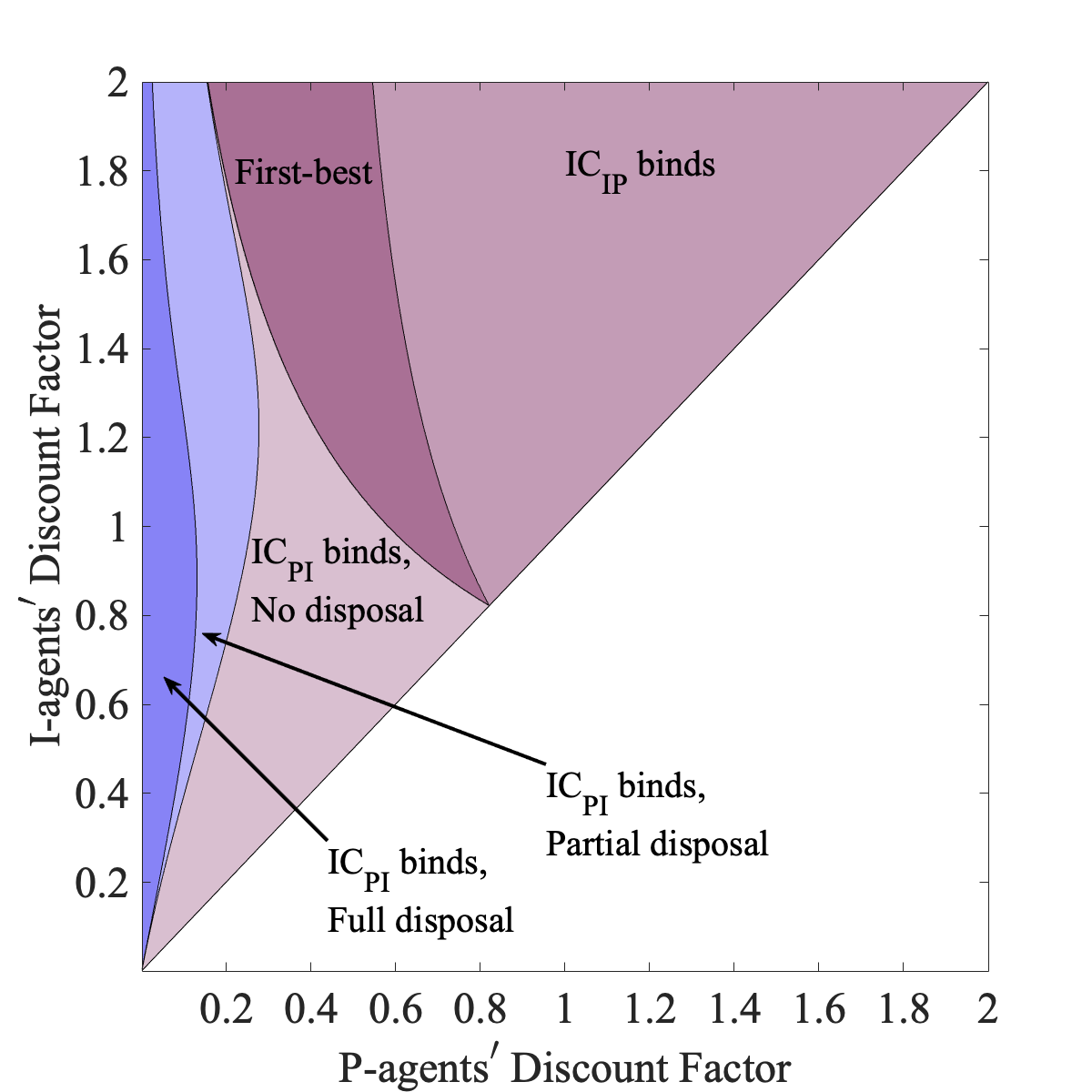}
	\caption{Second-best allocations in the timed homogeneous-good example.}
		\label{fig:SecondBestTypes}
\end{figure}

We illustrate the different regions corresponding to second-best allocations for our example of homogeneous goods that vary in their delivery times. Figure \ref{fig:SecondBestTypes} displays the structure of the second-best solution for different discounts (\textit{Full disposal} refers to solutions in which $I$-agents are either served with higher-quality goods than $P$-agents, or not served at all).\footnote{As before, the figure corresponds to an environment with an equal mass of each agent type: $\mu_I=\mu_P=1/2$, weighted equally by the mechanism designer. The supply has total mass of $3/2$ and is distributed uniformly over $[0,5]$.} Clearly, all allocation forms occur for a substantial set of parameters. Second-best allocations are also first-best allocations for intermediate values of discount factors. Finally, disposal occurs when $P$-agents' discount factor is low and there is a sufficient wedge between the discount factors.

\subsection{Welfare Implications}

We now discuss welfare properties of our second-best allocation. As a benchmark, we consider the pooling allocation, which is incentive compatible and inherently ``fair,'' but never optimal. Who gains and who loses as we move from the pooling allocation to the second-best allocation?

Welfare comparisons can be determined via which $IC$ constraint binds, using two observations. First, when $IC_{kj}$ binds, $k$-agents are indifferent between their allocation and $j$-agents' allocation. They are therefore indifferent between the allocation they receive and any \textit{mixture} of the two. When there is no disposal, they are then also indifferent between the allocation they receive and the pooling allocation. It follows that, when there is no disposal, if $IC_{kj}$ binds, $k$-agents are as well off as in the pooling allocation. With disposal, mixtures of the second-best allocations each type receives generate strictly lower utility than the pooling allocation.

Second, since the pooling allocation is never optimal, at least one of the agents must prefer the second-best. It follows that $k$-agents have strictly higher welfare in the second-best than in the pooling allocation if and only if $IC_{kj}$ does \textit{not} bind.

\begin{corollary}\label{cor:Welfare}
Suppose $(q_P, q_I)$ is a second-best solution. Then, one of the following must hold:
\begin{enumerate}
	\item Neither $IC_{IP}$ nor $IC_{PI}$ binds. Then, the solution coincides with the first-best solution. Both $I$- and $P$-agents strictly prefer it to the pooling allocation.
	\item Only $IC_{IP}$ binds. Then, $P$-agents strictly prefer the second-best to the pooling allocation, while $I$-agents are indifferent.
	\item Only $IC_{PI}$ binds and there is no disposal. $I$-agents strictly prefer the second-best to the pooling allocation, while $P$-agents are indifferent.
	\item Only $IC_{PI}$ binds and there is disposal. $I$-agents strictly prefer the second-best to the pooling allocation, while $P$-agents strictly prefer the pooling to the second-best allocation.
\end{enumerate}
\end{corollary}

\section{Beyond Two Types\label{sect:Ntypes}}

We now turn to discuss how our results extend to the case of any finite number $N$ of types. We maintain the same assumptions for the utility $u_i$ of each type $i$. As before, we posit that all types have the same ordinal preferences, but differ in their cardinal valuations, with utilities ordered via absolute risk aversion: $\forall x \in \mathbb{R}_+$,

\begin{equation}
\frac{u_{i}^{\prime \prime }(x)}{u_{i}^{\prime }(x)}>\frac{u_{i+1}^{\prime
\prime }(x)}{u_{i+1}^{\prime }(x)}\ \ \ \ 
\forall i \in \{1, \dots, N-1\} \text{.}  \label{eq:Utility_RankingNTypes}
\end{equation}

Each type $i$ has mass of $\mu_{i}>0$ and we continue to focus on a setting with sufficient supply: $\sum_{i=1}^{N}\mu_{i}\leq F(X)$.

\subsection{First-best with Many Types}

As in the two-type case, the social planner selects an allocation $(q_{1},...,q_{N})$, where each $q_{i}$ is a lottery and the allocation is feasible:
\begin{equation*}
\sum_{i=1}^{N}\mu_{i}q_{i}(x)\leq f(x)\text{ \ \ }\forall x\in\lbrack 0,X].
\end{equation*}
The goal is to maximize welfare
\begin{equation*}
W(q_{1},...,q_{n})=\sum_{i=1}^{N}\alpha_{i}\mu_{i}\int_{0}^{X}u_{i}(x)q_{i}(x)dx\text{,}
\end{equation*}%
where $\{\alpha_{i}\}_{i=1}^{n}$ are arbitrary weights with $\alpha_{i}>0$ for all $i$. 

Following our results for the two-type case, we can expect any first-best solution to satisfy two properties. First, it is never optimal for the planner to dispose of goods. Second, there are no inverted spreads: for any $i$ and $j$ with $i<j$, if any part of $i$-agents' allocation is a spread of any part of $j$-agents' allocation, there is a beneficial swap that violates optimality.

As it turns out, these two restrictions fully characterize the first-best: the set of allocations that exhibit no disposal and no inverted spread coincides with the set of first-best allocations for some welfare weights. Moreover, this set also coincides with the set of Pareto efficient allocations. 

\begin{proposition}
	The following sets coincide:
	\begin{enumerate}
    \item The set of feasible allocations that do not exhibit disposal or inverted spread;
    \item The set of feasible allocations that are first-best for some strictly positive welfare weights $\{\alpha_{i}\}_{i=1}^{N}$;
    \item The set of feasible Pareto efficient allocations.
    \end{enumerate}
\end{proposition}

To glean intuition, consider a market for allocations in which agents' initial allocations serve as their endowments and trade can take place freely---we further elaborate on such markets in the following section. From the second welfare theorem, any Pareto efficient allocation can be mapped to a competitive equilibrium for some endowments. An inverted spread would leave room for beneficial trades and cannot occur in a competitive equilibrium. Conversely, without an inverted spread, agents have no opportunities for profitable bilateral trade. As it turns out, there is also no profitable multilateral trade.\footnote{The proof itself uses alternative arguments.} An allocation without an inverted spread is then a competitive equilibrium for some endowments and thus Pareto efficient. Finally, standard arguments show the equivalence between Pareto efficiency and utilitarian efficiency with some welfare weights.

\paragraph{Shape of the first-best allocation} What do allocations that exhibit no inverted spread and no disposal look like? As in the $N=2$ case, we denote by $\overline{X}:=F^{-1}(\mu_1+...+\mu_N)$ the lowest quality needed to exhaust demand when only the best-quality goods are used. Any allocation satisfying no disposal fully utilizes goods in $[0,\overline{X}]$. 

The most risk-averse agents must be served in one contiguous block $[x_{1},x^{1}] \subseteq [0,\overline{X}]$, exhausting the supply available there; if another type were served within the block, it would yield an inverted spread. Thus, we have $q_{1} = f ~\vert~ [x_{1},x^{1}]$ with $f([x_{1},x^{1}]) = \mu_{1}$. Once we determine type-$1$ agents' allocation, we can consider a reduced problem. We adjust the supply by eliminating the block of goods already promised to type-$1$ agents. Namely, we let $f^1:=f -\mu_1 q_1$ and focus on agents of types $2,...,N$. Of those, type-$2$ agents are the most risk averse and, as before, must be served in one block, exhausting the supply $f^{1}$ there. Thus, $q_{2} = f^{1} ~\vert~ [x_{2},x^{2}]$ for some $0 \leq x_{2} < x^{2} \leq \overline{X}$. While $q_2$ exhausts the supply $f^1$ within $[x_{2},x^{2}]$, it need not be a continuous block within $[0,\overline{X}]$---we may have $[x_{1},x^{1}] \subset [x_{2},x^{2}]$. We can continue to generate assignments for all types recursively.\footnote{Namely, at step $k$, define the remaining supply $f^{k-1}=f - \sum_{i=1}^{k-1}\mu_{i}q_{i}$. Since there is no inverted spread, $q_{k} = f^{k-1} ~\vert~ [x_{k},x^{k}]$ for some $0 \leq x_{k} < x^{k} \leq \overline{X}$. Without loss of generality, we assume that $x_{k}=\textrm{inf}(\textrm{supp}(q_{k}))$ and $x^{k}=\textrm{sup}(\textrm{supp}(q_{k}))$. At each step $k$, the pair $x_{k},x^{k}$ respects a feasibility constraint $f^{k-1}([x_{k},x^{k}])=\mu_{k}$.} Denote by $\mathcal{A}$ the set of all allocations that can be constructed using this procedure. It turns out that all first-best allocations can be constructed in this way and that any such construction leads to a first-best allocation.

\begin{corollary}
	The set $\mathcal{A}$ of allocations coincides with the set of allocations that do not exhibit disposal or inverted spread. In particular, any first-best allocation consists of no more than $2N-1$ blocks and type-$k$'s allocation consists of no more than $k$ disjoint blocks.
\end{corollary}

\subsection{Many Unobservable Types}

The mechanism designer's problem is defined analogously to that for $N=2$: the designer chooses feasible lotteries to maximize the weighted sum of utilities subject to incentive-compatibility constraints $IC_{kj}$ with $k\neq j$ and $k,j=1,...,N$, ensuring that $k$-agents do not want to emulate $j$-agents.

To derive properties of the solution, we further assume that utility functions of different types are linearly independent. Formally, for any $\lambda_{0},...,\lambda _{N}\in \mathbb{R}$, the set $\left\{~ x~ \big \vert~ \sum_{j=1}^{N}\lambda_{j}u_{j}(x)=\lambda_{0}~\right\}$ has measure zero. This assumption, which holds automatically with only two types, is valid for many classes of utilities, including CRRA, CARA, exponentially discounted utilities, or present-biased ones. This regularity assumption guarantees that a social planner is never indifferent between randomly supplying an interval of goods to several types, or assigning those goods to another, different type.

\begin{proposition}\label{th:NTypes}
	A solution of the mechanism designer's problem with $N$ types exists and is unique. If $(q_i)_{i\in \{1, \dots, N\}}$ is a solution,
	\begin{enumerate}
		\item For almost every $x \in [0, X]$, either one type of agent gets the entire supply of the good or the entire supply remains unused. That is, for almost every $x \in [0, X]$, either $q_i(x)=f(x)$ or $q_i(x)=0$ for all $i\in \{1, \dots, N\}$. Thus, the solution is fully separating.
		\item The graph of binding IC constraints for the optimal allocation has no directed cycles. That is, there is no subset of types $(k_1, \dots, k_m) \subset N^m$, with $k_i \neq k_j$ for all $i, j \in \{1, \dots, m\}$  such that 
		$IC_{k_{i}, k_{i+1}}$ for all $i\in \{1, \dots,  m-1\}$ and $IC_{k_{m}, k_{1}}$ all bind.
	\end{enumerate}
\end{proposition}

The proposition indicates that some of the main properties of our solution with two types continue to hold. A solution exists and is unique. Part 1 of the proposition illustrates that it is ``fully'' separating---not only do different types get different allocations, but their allocations' supports do not meaningfully overlap. This implies, once again, that the pooling allocation is never a solution and yields strictly lower welfare.  

Part 2 of the proposition asserts that the $IC$ constraints do not form a cycle, generalizing the observation from our two-type setting that both $IC$ constraints cannot bind simultaneously. It implies that the $IC$ constraints cannot all be binding even in the $N$-type case. 

Recall our Corollary 4 that suggested that any allocation in the set $\mathcal{A}$ is a first-best allocation for some welfare weights. In general, to establish incentive compatibility, there are $N \times (N-1)$ constraints that need to be satisfied. However, by construction, for any allocation in $\mathcal{A}$, it suffices to check the constraints corresponding to adjacent types---$k$-agents' allocation contains more extreme-quality goods than $k-1$-agents'. This insight allows us to illustrate the possible coincidence of the first-best and second-best allocations for a non-trivial set of welfare weights. 

Intuitively, construct an allocation in $\mathcal{A}$ as follows. Pick some $[x_{1},x^{1}] \subset (0,\overline{x})$. We can find an allocation for type-$2$ agents, defined by $y_{2}$ and $y^{2}$ so that type-$1$ agents are indifferent between their allocation and the resulting type-$2$ agents' allocation.\footnote{As in the construction of the set $\mathcal{A}$, type-$2$ agents are provided goods of quality $[y_{2},x_{1}] \cup [x^{1},y^{2}]$.} Since type-$2$ agents are less risk averse, they strictly prefer their allocation. We then find an allocation for type-$2$ agents, defined similarly by $z_{2}$ and $z^{2}$, such that type-$2$ agents are indifferent between their allocation and type-$1$ agents' allocation. It follows that type-$1$ agents would strictly prefer their allocation. Consider now the allocation defined by $x_{2}=\frac {y_{2}+z_{2}}{2}$ and $x^{2}=\frac{y^{2}+z^{2}}{2}$, assuming it is feasible. Our construction guarantees that neither $IC_{12}$ nor $IC_{21}$ binds. We can complete the construction of such an allocation for higher types, each step ensuring that incentive compatibility constraints of adjacent types do not bind (in the proof, we make appropriate adjustments to guarantee feasibility). By Corollary 4 and Proposition 4, this allocation is a first-best solution for some welfare weights. Since it is incentive compatible, under those same welfare weights, it is also a second-best solution. Furthermore, any small enough perturbation of the welfare weights corresponds to a small perturbation of the first-best solution---namely, a small change in the end-points defining the allocations of each agent type---and remains incentive compatible. 

\begin{corollary}
For all $N>2$, there is an open set of welfare weights for which the first-best allocation is incentive compatible.
\end{corollary}

To illustrate graphically the welfare implications of the first- and second-best allocations for different type volumes, we focus on the special case of time discounting. We consider random profiles of $N$ discount factors, where $N=2,...,10$ and each discount factor is randomly and drawn uniformly from $[0,2]$. We assume all types have equal masses, supply is uniform, and all types are weighted equally in the expected welfare.\footnote{For any $N$, $\alpha_{i}=1$, $\mu_{i}=\frac{1}{N}$ for all $i \in \{1,...,N \}$. We simulate $100,000$ discount rates within $[0,2]$ and partition those randomly to sets of $N$. The supply has total mass $3/2$ and is distributed uniformly over $[0,5]$. }

\begin{figure}
	\centering
	\includegraphics[height=7cm]{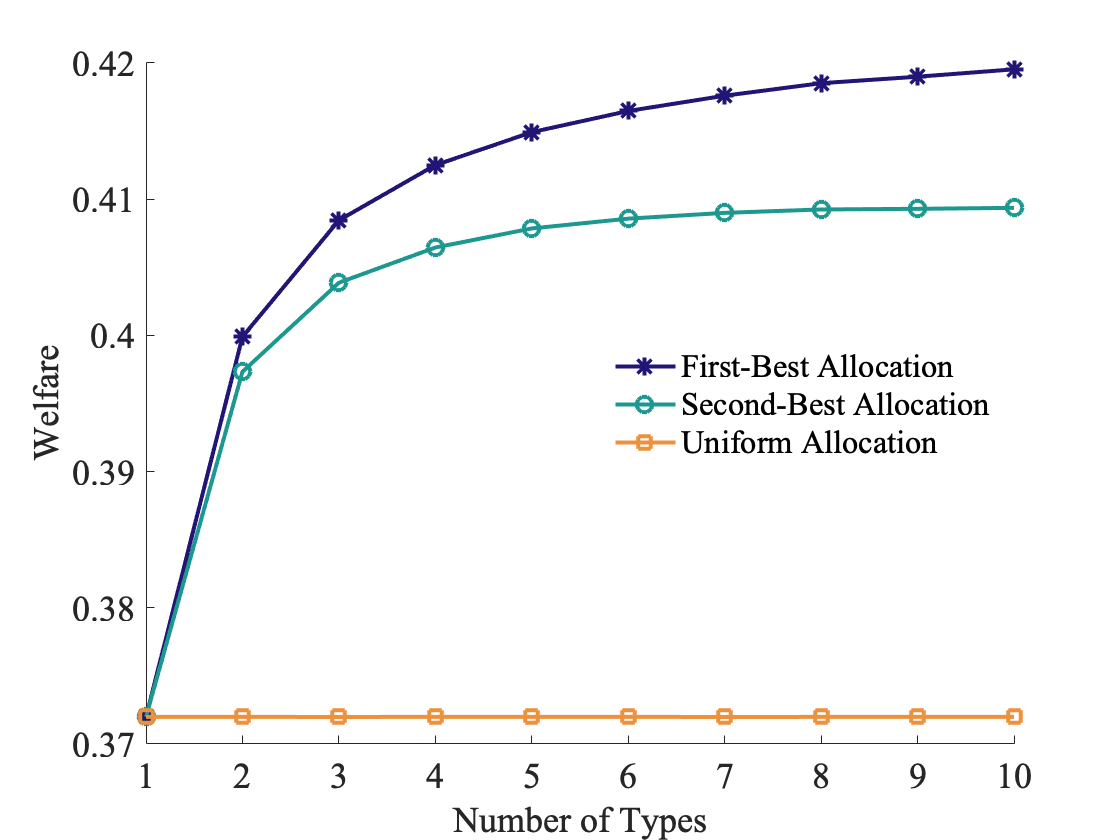}
	\caption{Welfare levels for $N$ types.}
		\label{fig:welfare}
\end{figure}

Figure \ref{fig:welfare} displays the expected welfare from the first-best, second-best, and pooling allocations. Naturally, the welfare from the first-best allocation exceeds that from the second-best allocation, and is lowest for the pooling allocation. As the figure illustrates, the expected welfare values plateau as the number of types increase. Furthermore, the second-best allocation establishes substantially higher welfare than the pooling allocation. 

\section{A Market for Allocations}

We now consider a simple method for generating incentive-compatible allocations that Pareto dominate the pooling allocation for an arbitrary number of types: a market for allocations, in the spirit of \cite{HyllandZeckhauser1979}. Agents receive an endowment and can trade freely through market interactions, which determine the prices of various allocations. As usual, such market interactions need not take place literally---they can be emulated after types are reported. We maintain the previous section's assumptions on agents' utilities.

Without loss of generality, we consider a symmetric notion of competitive equilibrium, where all agents of a certain type have the same demand for lotteries.\footnote{This is without loss of generality since, for any asymmetric equilibrium, we can construct a corresponding symmetric equilibrium with the same aggregate demand per type and the same price schedule.} A \textit{price schedule} is a measurable function $p:~[0,\overline{X}] \rightarrow R_{+}$. A \textit{demand} of type-$i$ agent is a measure $q_{i}$ over $[0,\overline{X}]$.\footnote{In principle, unlike lotteries described in prior sections, a demand function can exhibit mass points.} 

An agent of type $i$ optimizes the allocation subject to a budget constraint, determined by the endowment $\omega_{i}$ and the price schedule $p(x)$: the agent's problem is
\begin{equation} \label{Agent_problem_competitive_N}
\underset{q_{i} 
}{\textrm{max}}~ \int_{0}^{\overline{X}}  u_{i}(x)q_{i}(x)dx  \end{equation}
such that
\begin{equation}
~~~~~~1 - q_{i}\big([0,\overline{X}]\big)~\geq~ 0 
~~~~~~~~~\mathrm{and}~~~~~~~~
\omega_{i} - \int_{0}^{\overline{X}} p(x) dq_{i}(x) ~\geq~ 0. 
\notag
\end{equation}

A natural, seemingly equitable case to consider entails all agents receiving equal endowments. For any given profile of prices, it is tantamount to giving all agents an equal share of the supply, essentially their pooling allocation.\footnote{Targeted endowments would introduce incentive problems at the endowment-distribution stage. One could also consider random, but unequal endowments. Our main insights carry over.} We call the resulting allocation a \textit{fair competitive equilibrium}.

\medskip
\noindent \textbf{Definition.} A \textit{fair competitive equilibrium} consists of a price schedule $p(x)$ and demand functions $\{ q^{i}\}_{i =1,...,N}$ such that
\begin{enumerate}
\item $q_{i}$ solves (\ref{Agent_problem_competitive_N}) for each type of agent $i$.
\item All agents receive the same strictly-positive endowment, $\omega_{i} ~=~\omega$.
\item Market clearing holds, $\sum_{i=1}^{N}\mu_{i} q_{i}(x)  ~=~ f(x) \cdot \mathbbm{1}\big\{x \in [0,\overline{X}] \big\} $. 
\end{enumerate}

The following proposition characterizes the structure of fair equilibrium outcomes.

\begin{proposition}\label{th:MarketStructure}
A fair competitive equilibrium exists. If $\big(p,\{q_{i}\}_{i=1,...,N}\big)$ is a fair competitive equilibrium, there are threshold qualities $0=\underline{x}_{N} < \underline{x}_{N-1}<...<\underline{x}_{2}<\underline{x}_{1}< \overline{x}_{1}<\overline{x}_{2}<....<\overline{x}_{N-1}<\overline{x}_{N}=\overline{X}$ such that 
\begin{equation}
q_{k} ~~=~~ f ~\big\vert~  [\underline{x}_{k},\underline{x}_{k-1}] \cup [\overline{x}_{k-1},\overline{x}_{k}]~~~\textrm{for}~k>1,
~~and~~
q_{1} ~~=~~ f ~\big\vert~ [\underline{x}_{1},\overline{x}_{1}].
\notag
\end{equation}
\end{proposition}

The proposition demonstrates that the structure of a fair competitive equilibrium extends the IPI structure we obtained as a second-best solution for two types. The most risk-averse agents receive goods in one contiguous interval; the second most risk-averse agents receive goods in two surrounding intervals; and so on, with the least risk-averse agents receiving either the best or the worst goods. The resulting allocation belongs to $\mathcal{A}$ and, as Corollary 4 indicates, does not exhibit disposal or inverted spread.

From the first welfare theorem, we know the fair competitive equilibrium allocation is Pareto efficient. In particular, it dominates the pooling allocation for \textit{all} agents. Furthermore, by construction, it satisfies all agents' incentive compatibility constraints. Nevertheless, it need not coincide with the second-best solution. Furthermore, while the second-best solution may entail disposal of goods, the fair competitive equilibrium never does.

To illustrate the wedge between the market and second-best solutions, consider again the discounting case with two types. Standard techniques can be used to show that the fair competitive equilibrium is unique. 

\begin{figure}
	\centering
	\begin{subfigure}[t]{.49\textwidth}
\centering
	\includegraphics[width=\linewidth]{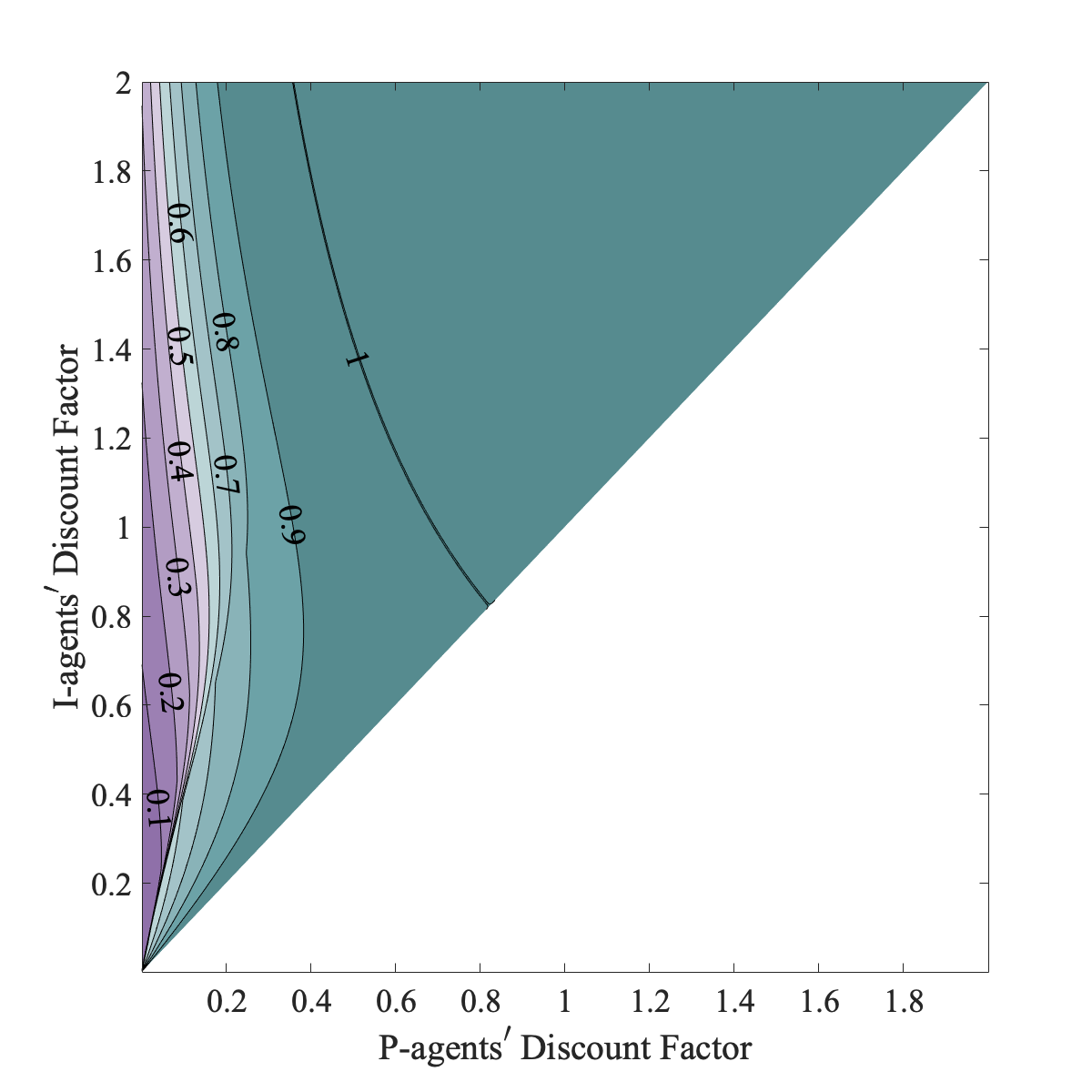}
\end{subfigure}
\begin{subfigure}[t]{.49\textwidth}
	\includegraphics[width=\linewidth]{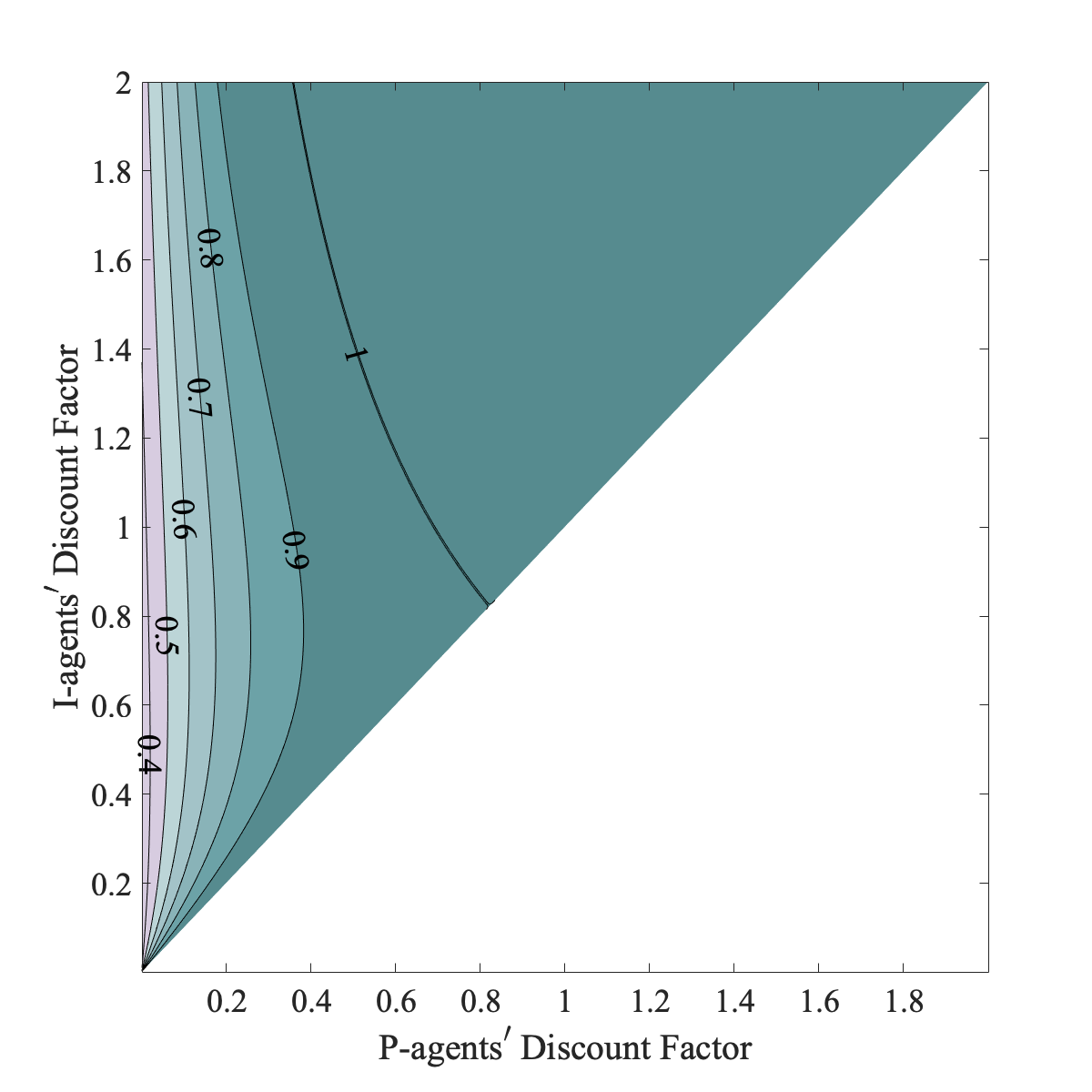}
\end{subfigure}
	\caption{Ratio of welfare generated by the fair competitive equilibrium relative to the second-best allocation, adjusted by welfare generated by the pooling allocation, with disposal (in left panel) and without (in right panel).}
		\label{fig:CEvsSBvsTB}
\end{figure}

Panel (a) of Figure \ref{fig:CEvsSBvsTB} displays the ratio of welfare generated by the fair competitive equilibrium and the second-best allocation, adjusted by welfare generated by the pooling allocation. There is a zero-measure set of discount rates---corresponding to the line with a ratio of $1$---for which there is no welfare loss produced by the market solution. For all other parameters, the second-best solution yields greater welfare levels. Welfare losses are more pronounced for more patient $P$-agents. Certainly, these discount-rate regions may entail disposal in the second-best allocation, which is never featured in the fair competitive equilibrium. Nonetheless, the wedge in welfare is not solely due to disposal: panel (b) of Figure \ref{fig:CEvsSBvsTB} considers restricted second-best solutions in which disposal is banned (see the Online Appendix for a formal analysis). Welfare losses generated by the market solution are still pronounced, particularly when $P$-agents are very patient.     

\section{Conclusions}\label{sect:Variants}

Goods and services---public housing, medical appointments, schools---are often allocated to individuals who rank them similarly but differ in their preference intensities. We characterize optimal allocation rules in such settings, considering both the case in which individual preferences are known and ones in which they need to be elicited. We show that first-best allocations may involve assigning some agents lotteries between high- and low-ranked goods. When preference intensities are private information, second-best allocations always involve such lotteries and may coincide with first-best allocations. Furthermore, second-best allocations may entail disposal of services. We also illustrate the potential drawbacks of utilizing a simple market solution in lieu of the optimal mechanism.

Our analysis assumes a fixed supply that cannot be altered, but complete freedom in selecting which agents are offered a good. In many applications, however, goods' quality can be reduced and denial of service is not viable. In the Online Appendix, we show that the possibility to reduce goods' qualities would never be utilized in the first- and second-best solutions. Furthermore, the qualitative features of the second-best solution are retained when all agents need to be catered to with certainty.

\appendix

\section*{\huge Appendix}


\begin{description}
\item[Proposition 0 (Existence)] \textit{For any number $N$ of types, a first-best and a second-best allocation exists.}
\end{description}

\noindent \textbf{Proof of Proposition 0}.
Since $Y=[0,X]$ is a compact metric space, the space of distributions $\mathcal{P}(Y)$ is a metric space. Moreover, it is a closed subset of the unit ball with respect to the weak$^*$ topology. The latter is a compact set by the Banach-Alaogly Theorem, hence $\mathcal{P}(Y)$ is a compact metric space. The space of allocations is a direct product $\left(\mathcal{P}(Y)\right)^{N}$, where $N$ is the number of agent types. Therefore, the space of allocations is a compact metric space with the $sup$ metric induced by the metric on $\mathcal{P}(Y)$. The $IC$ and feasibility constraints are not strict and are linear in the allocation. Hence, the subset of feasible and incentive-compatible allocations is a closed subset of $\left(\mathcal{P}(Y)\right)^{N}$, and in itself a compact set. Finally, the objective function (of both the social planner and the mechanism designer) is linear in the allocation, and, therefore, continuous. A first-best and second-best allocation exists by the Weierstrass Theorem.\hfill $\blacksquare$

\begin{description}
\item[Lemma 0] \textit{Agents of type $P$ are strictly more risk averse than agents of type $I$.}
\end{description}
\noindent \textbf{Proof of Lemma 0.} The result is well known for lotteries with support on $[0,\infty)$. For a lottery with $\textrm{supp}(q) \subseteq [0,X] \cup \{\diamond\}$ and $q_{\diamond}>0$, there is a sequence of lotteries $q^{(k)} = \big(1-q(\diamond)\big) \cdot q ~\vert~ [0,X] +q(\diamond) \cdot \delta_{k}$, $k=1,2,...$, defined on the Borel subsets of $[0,\infty)$, such that $V_{j}(q^{(k)}) \overset{k \rightarrow \infty}{\longrightarrow} V_{j}(q)$. Thus, we can use known results for lotteries with support on $[0,\infty)$ to conclude that $P$-agents are weakly more risk averse than $I$-agents. The proof that this ordering is, in fact, strict for lotteries with $q(\diamond)>0$ is also standard, but somewhat more involved: it appears as Lemma A1 in the Online Appendix.\hfill $\blacksquare$

\noindent \textbf{Proof of Lemma 1.}
We are interested in the sign of the derivative of $g(\cdot)$:
\begin{equation}
g'(x) ~=~ \alpha u_{P}'(x) - (1-\alpha) u_{I}'(x)
\notag
\end{equation}
Since utilities are decreasing,
\begin{equation}
\textrm{sign} (g'(x)) ~=~ 
-\textrm{sign} \left( \frac{u_{P}'(x)}{u_{I}'(x)} ~-~ \frac{1-\alpha}{\alpha} \right) ~=~ - \textrm{sign} (\theta(x)),
\notag
\end{equation}
where $\theta(x) = \frac{u_{P}'(x)}{u_{I}'(x)} - \frac{1-\alpha}{\alpha}$.
We have
\begin{equation}
\theta'(x) ~=~ \frac{u_{P}'(x)}{u_{I}'(x)} \cdot \left[ \frac{u_{P}''(x)}{u_{P}'(x)} ~-~ \frac{u_{I}''(x)}{u_{I}'(x)}\right] ~>~0.
\notag
\end{equation}
Thus, $\theta$ is a strictly increasing function. There are three possible cases. If $\textrm{sign}(\theta(x))>0$ always, then $g(\cdot)$ is strictly decreasing and, hence, strictly quasi-concave. If  $\textrm{sign}(\theta(x))<0$ always, then $g(\cdot)$ is strictly increasing and, hence, strictly quasi-concave. Last, suppose there is $x_{PI}$ such that $\textrm{sign}(\theta(x))<0$ for $x<x_{PI}$, $\textrm{sign}(\theta(x))>0$ for $x>x_{PI}$, and $\textrm{sign}(\theta(x_{PI}))=0$. Consider $x'<x''$ and $\lambda \in (0,1)$. If $x' < \lambda x' +(1-\lambda)x'' \leq x_{PI}$, then $g(\lambda x' + (1-\lambda )x'') > g(x')$. Otherwise, $x_{PI} \leq \lambda x'+(1-\lambda)x'' < x''$ and $g(\lambda x' + (1-\lambda)x'') > g(x'')$. It follows that $g$ is a strictly quasi-concave function, since $g(\lambda x' +(1-\lambda)x'') > \textrm{min}\{g(x'),g(x'')\}$ for all $\lambda \in (0,1)$. \hfill $\blacksquare$ 

\vspace{3mm}

\noindent \textbf{Proof of Proposition 1}
An optimal allocation exists by Proposition 0. If $q$ is an optimal allocation, then $(\mu_{I}q_{I}+\mu_{P}q_{P})([0,\overline{X}])=\mu_{I}+\mu_{P}=F(\overline{X})$. Otherwise, an allocation for at least one of the agents' types could be improved through the provision of superior-quality goods instead of goods on $(\overline{X},\infty) \cup \{\diamond\}$, without altering the other type agents' allocation.
Therefore, if $q_{P}$ is chosen optimally, we can write $q_{I}=q_{I}(q_{P})$, where $q_{I}(x) = \mu_{I}^{-1}\cdot(f(x)-\mu_{P}q_{P}(x)) \cdot \mathbbm{1}\{ x \in [0,\overline{X}] \}$. The social planner's optimization problem is then: 
\begin{equation}
\max_{q_{P}}~~(1-\alpha)\int_{0}^{\overline{X}}f(x)u_{I}(x)dx+ \mu_{p} \int_{0}^{\overline{X}} g(x)q_{P}(x)dx
\notag
\end{equation}
\begin{equation}
\hbox{s.t.:}~~~~ 0 \leq q_{P}(x) \leq \mu_{P}^{-1}f(x) ~~~~~~,~~~~~~ \int_{0}^{\overline{X}}q_{P}(x)dx =1
\notag
\end{equation}
This problem has a unique solution given by $q_{p}(x) = \mu_{P}^{-1}f(x) \mathbbm{1} \{ g(x) \geq c \}$, where $c = \inf \big\{ t \in R ~\big\vert~ \mu_{P}^{-1}f(\{ x \in [0,\overline{X}] ~\vert~ g(x)>t\}) \leq 1 \big\}$ by the so-called ``Bathtub'' principle (Theorem 1.14 in \cite{LiebLossAnalysis}). The arguments provided in Section 3 show that the solution has the desired form in all three cases considered in the proposition. To see that the 3 cases are not overlapping, assume that $g(\overline{X}_{P}) \leq g(0)$. The strict quasi-concavity of $g(\cdot)$ implies that $\min \{g(0),g(\overline{X}) \} < g(\overline{X}_{P}) \leq g(\overline{X}_{P})$. Therefore, $g(\overline{X})=\min \{g(0),g(\overline{X}) \} <g(\overline{X}_{I})$. \hfill $\blacksquare$ 

The following lemma generalizes Lemma 2 in the text and will be used in our $N$-type analysis.

\vspace{3mm}

\noindent \textbf{Lemma 2* (No Inverted Spread).} \textit{The first-best allocation never exhibits an inverted spread for any number of types. The second-best allocation exhibits no inverted spread for N=2 types.}

\vspace{3mm}

\noindent \textbf{Proof of Lemma 2*.} 
Consider first the case of $N=2$ types. Suppose the allocation $(q_{P},q_{I})$
exhibits an inverted spread; that is, there are $A \triangleleft B \triangleleft C$ such that $q_{P}(A),q_{I}(B),q_{P}(C)>0$. For an arbitrary $x \in B$, let $\gamma(x) \in (0,1)$ be such that $V_{P}\big(\gamma(x) \cdot q_{P} ~\vert~A +(1-\gamma(x)) \cdot q_{P} ~\vert~ C\big) = V_{P}\big(\delta_{x}\big)$, then  $V_{I}\big(\gamma(x) \cdot q_{P} ~\vert~A +(1-\gamma(x)) \cdot q_{P} ~\vert~ C\big) > V_{I}\big(\delta_{x}\big)$ by Lemma 0. Integrating these inequalities with respect to $q_{I} ~\vert~B$ and using the linearity of $V_{P}(\cdot),V_{I}(\cdot)$, we get
\begin{equation}
V_{P}\big(\gamma \cdot q_{P} ~\vert~ A + (1-\gamma) \cdot q_{P} ~\vert~ C\big) = V_{P}\big(q_{I} ~\vert~ B\big) ~~~~~~~,~~~~~~~
V_{I}\big(\gamma \cdot q_{P} ~\vert~ A + (1-\gamma) \cdot q_{P} ~\vert~ C\big) > V_{I}\big(q_{I} ~\vert~ B \big)
\notag
\end{equation}
where $\gamma = \int \gamma(x) dq \vert B (x) \in (0,1)$. Let $\epsilon = \textrm{min} \{q_{P}(A),q_{I}(B),q_{P}(C) \}$, then $\epsilon>0$. Consider
\begin{equation}
q'_{I} ~~=~~ q_{I} ~-~ \epsilon \mu_{P}(\mu_{P}+\mu_{I})^{-1}  \cdot q_{I} ~\vert~ B ~+~ 
\epsilon \mu_{P}(\mu_{P}+\mu_{I})^{-1}  \cdot \big(\gamma\cdot q_{P} ~\vert~ A + (1-\gamma)  \cdot q_{P} ~\vert~ C \big)
\notag
\end{equation}
\begin{equation}
q'_{P} ~~=~~ q_{P} ~+~ \epsilon \mu_{I}(\mu_{P}+\mu_{I})^{-1}  \cdot q_{I} ~\vert~ B ~-~ 
\epsilon \mu_{I}(\mu_{P}+\mu_{I})^{-1}  \cdot \big(\gamma \cdot q_{P} ~\vert~ A + (1-\gamma)  \cdot q_{P} ~\vert~ C \big)
\notag
\end{equation}
It is easy to see that $q'$ is feasible. Since $V_{P}(q'_{P})=V_{P}(q'_{P})$, $V_{I}(q'_{I})>V_{I}(q_{I})$, then $W(q')>W(q)$. Thus, $q$ cannot be a first-best allocation. Notice also that $V_{P}(q'_{I})=V_{P}(q_{I})$, and $V_{I}(q'_{P})<V_{I}(q_{I})$. Therefore, if $q$ is incentive compatible, then $q'$ is also incentive compatible. We conclude that $q$ cannot be a second-best allocation as well.

Suppose that $N>2$ and types $j,k$ exhibit an inverted spread, where type $j$ is more patient than type $k$. Fix an allocation for all types $i \neq j,k$ and repeat the argument used for $N=2$ types for $j=P$ and $k=I$ to conclude that the first-best allocation cannot exhibit an inverted spread.\footnote{Our argument for the second-best allocation does not extend beyond two types. Indeed, using the notation above, some of the incentive constraints $IC_{ij},IC_{ik}$ for $i \neq j,k$ may be violated for $q'$.} \hfill $\blacksquare$

\vspace{3mm}

\noindent \textbf{Proof of Corollary 2.}
Let $q$ be an optimal allocation. First, let $\nu(\cdot)$ be a $\mathcal{L}$ebesque measure, and assume that $\nu(\textrm{supp}(q_{I}) \cap \textrm{supp}(q_{P}) \cap [0,X]) >0.$ Since $q_{P},q_{I}$ are non-atomic, there are $0<x'<x''<X$ such that $\nu(\textrm{supp}(q_{I}) \cap \textrm{supp}(q_{P}) \cup [0,x'])>0$, $\nu(\textrm{supp}(q_{I}) \cap \textrm{supp}(q_{P}) \cup (x',x''))>0$, and $\nu(\textrm{supp}(q_{I}) \cap \textrm{supp}(q_{P}) \cup [x'',X])>0$. Then $[0,x'] \triangleleft (x',x'') \triangleleft [x'',X]$ constitutes an inverted spread, contradicting Lemma 2.

Furthermore, if $q_{j}(\diamond)=1$ for some type, then by incentive compatibility $q_{P}=q_{I}=\delta_{\diamond}$, which clearly cannot be optimal. Hence, $q_{I}(\diamond),q_{P}(\diamond)<1$. 
Assume, towards a contradiction, that both $IC$ constraints are binding. Then the allocation $\widetilde{q}_{P}=\widetilde{q}_{P} = \dfrac{\mu_{P}q_{P}+\mu_{I}q_{I}}{\mu_{P}+\mu_{I}}$ is also feasible, incentive compatible, and provides the same welfare as $q$. It follows that this allocation is also optimal. The supports of $\widetilde{q}_{I}$ and $\widetilde{q}_{P}$ coincide, and $\widetilde{q}_{I}(\diamond),\widetilde{q}_{P}(\diamond) <1$. These observations imply that  $\nu \big(\textrm{supp}(\widetilde{q}_{I}) \cap \textrm{supp}(\widetilde{q}_{P}) \cap [0,X] \big) >0$, in contradiction. \hfill $\blacksquare$

\vspace{3mm}

\noindent \textbf{Proof of Proposition 2.}
Define $x_{2}: [0,F^{-1}(\mu_{I})] \rightarrow [F^{-1}(\mu_{P}),\overline{X}]$ as $x_{2}(x_{1}) = F^{-1}(F(x_{1})+\mu_{P})$, 
\begin{equation}
    \underline{x}_{1} ~~=~~ \min \Big\{ x_{1} \in [0,F^{-1}(\mu_{I})] ~~\Big \vert~~ V_{I}\big(f ~\big \vert~ [0,x_{1}] \cup [x_{2}(x_{1}),\overline{X}]\big) \geq
    V_{I}\big(f ~\big \vert~ [x_{1},x_{2}(x_{1})]\big)
    \Big\},
    \notag
\end{equation}
\begin{equation}
    \overline{x}_{1} ~~=~~ \max \Big\{ x_{1} \in [0,F^{-1}(\mu_{I})] ~~\Big \vert~~ V_{P}\big(f ~\big \vert~ [0,x_{1}] \cup [x_{2}(x_{1}),\overline{X}]\big) \leq
    V_{P}\big(f ~\big \vert~ [x_{1},x_{2}(x_{1})]\big)
    \Big\}.
    \notag
\end{equation}
Thus, if $x_{1} = \underline{x}$, then $I$-agents are indifferent between lottery $f ~\big \vert~ [0,x_{1}] \cup [x_{2}(x_{1}),\overline{X}]$ and lottery $f ~\big \vert~ [x_{1},x_{2}(x_{1})]$. It follows that the more risk averse $P$-agents strictly prefer the second lottery. Therefore, $\overline{x}_{1}>\underline{x}_{1}$. Similarly, if $x_{1} = \overline{x}_{1}$, $P$-agents are indifferent between the two lotteries above. In this case, $I$-agents strictly prefer the first lottery. By construction, for any $x_{1} \in [\underline{x}_{1},\overline{x}_{1}]$, the allocation $q^{x_{1}}$ given by $q_{P}^{x_{1}} = f ~\big \vert~ [x_{1},x_{2}(x_{1})]$, $q_{I}^{x_{1}} = f ~\big \vert~ [0,x_{1}] \cup [x_{2}(x_{1}),\overline{X}]$ is feasible and incentive compatible. It follows that $0<\underline{x}_{1} < \overline{x}_{1} < F^{-1}(\mu_{I})$.

By Proposition 1, for any $\alpha \in (0,1)$, the first-best allocation takes the form $q^{x_{1}}$, as defined above, for some $x_{1} \in [0,F^{-1}(\mu_{I})]$. Therefore, we can define a function $x_{1}: (0,1) \rightarrow [0,F^{-1}(\mu_{I})]$ by identifying $x_{1}(\alpha)$ such that $q^{x_{1}(\alpha)}$ is the unique first-best allocation for welfare weight $\alpha$.
For brevity, in what follows, we drop the arguments of $x_{1}$ and $x_{2}$ whenever there is little risk of confusion. Suppose $x_{1} \in (0,F^{-1}(\mu_{I}))$. By Proposition 1, $g(x_{1})=g(x_{2})$. There is therefore an inverse function $\alpha: (0,F^{-1}(\mu_{I})) \rightarrow (0,1)$ given by \begin{equation*}
\alpha(x_{1}) = \dfrac{u_{I}(x_{1})-u_{I}(x_{2})}{u_{P}(x_{1})-u_{P}(x_{2})+ u_{I}(x_{1})-u_{I}(x_{2})} = \dfrac{1}{1+\gamma(x_{1})},    
\end{equation*}
where 
\begin{equation*}
\gamma(x_{1}) = \dfrac{u_{P}(x_{1})-u_{P}(x_{2})}{u_{I}(x_{1})-u_{I}(x_{2})}.
\end{equation*}

Our assumptions on utilities guarantee that $\gamma$ is strictly increasing: see Lemma A2 in the Online Appendix for a complete argument. Therefore, $\alpha(\cdot)$ is strictly decreasing. It follows that the first-best allocation is incentive compatible if and only if $\alpha \in [\alpha(\underline{x}_{1}),\alpha(\overline{x}_{1})]$. \hfill $\blacksquare$

\noindent \textbf{Proof of Lemma 3.}
Assume that $IC_{ij}$ does not bind for $i \in \{I,P \}$, $j \in \{I,P \} \backslash i$. Then, $q_{j}(\diamond)=0$. To see this, suppose $q_{j}(\diamond) > 0$. Since the supply is sufficient, $(f-\mu_{P}q_{P}-\mu_{I}q_{I})([0,X])>0$. Then for sufficiently small $\epsilon>0$ the allocation $q'$ with $q'_{i}=q_{i}$ and
\begin{equation}
 q'_{j}(x) ~~~~~=~~~ 
 \Big(q_{j}(x) + \epsilon \cdot q_{j}(\diamond) \cdot \big(f(x)-\mu_{P}q_{P}(x)-\mu_{I} q_{I}(x)\big) \Big) \cdot \mathbbm{1} \big\{ x \in [0,X] \big\} ~+~
 \notag
 \end{equation}
 \begin{equation}
~+~ \Big(1-\epsilon \cdot (f-\mu_{P}q_{P}-\mu_{I}q_{I})\big([0,X]\big) \Big) \cdot q_{j}(\diamond) \cdot \delta_{\diamond}(x)
 \notag
 \end{equation}
is feasible, incentive compatible, and provides a strict welfare improvement with respect to $q$, thereby producing a contradiction.

We now show that $q_{P} = f ~\vert~ [x_{1},x_{2}]$ with $F(x_{2})-F(x_{1}) = \mu_{P}$. By Corollary 2, both $IC$ constraints cannot bind for $q$.

There are two cases to consider. First, suppose $IC_{IP}$ is not binding. Our argument above shows that $q_{P}(\diamond) = 0$. Therefore, we can define $x_{1} = \inf (\textrm{supp}(q_{P}))$, $x_{2} = \sup (\textrm{supp}(q_{P}))$, with $x_{1},x_{2} \in [0,X]$. By Corollory 2, $\nu( \textrm{supp}(q_{I}) \cap (x_{1},x_{2}) )=0$, and hence $q_{I}([x_{1},x_{2}]) = 0$. Assume, towards a contradiction, that $(f-\mu_{p}q_{p})([x_{1},x_{2}]) >0$. Let $x'_{2} = F^{-1}\big(F(x_{1})+\mu_{P}\big)$. Then $x'_{2}<x_{2}$ and $q_{P}\big((x'_{2},x_{2}]\big)>0$. Consider allocation $q'$ with $q'_{I}=q_{I}$ and 
\begin{equation}
q'_{P}(x) ~~~~~=~~~  \Big(q_{P}(x) + \epsilon \cdot q_{P}\big([x_{2}',x_{2}]\big) \cdot \big(f(x)-\mu_{P}q_{P}(x)\big) \Big) \cdot \mathbbm{1} \big\{ x \in [x_{1},x'_{2}] \big\} ~+~
\notag 
\end{equation}
\begin{equation}
~+~ \Big(1-\epsilon \cdot (f-\mu_{P}q_{P})\big([x_{1},x'_{2}]\big) \Big) \cdot q_{P}(x) \cdot \mathbbm{1} \big\{ x \in (x'_{2},x_{2}] \big\}
 \notag
 \end{equation}
For small enough $\epsilon>0$, the allocation $q'$ is feasible, incentive compatible, and provides a strict welfare improvement with respect to $q$, in contradiction.
 
The second case to consider corresponds to $IC_{IP}$ binding, in which case $IC_{PI}$ does not bind. As above, $q_{I}(\diamond) =0$. Using our definitions of $x_{1}$, $x_{2}$, and $x'_{2}$, Corollary 2 implies that $q_{I}([x_{1},x_{2}])$=0. If $q_{I}\big([x_{2},X] \big)=0$, then $V_{P}(q_{I}) \geq u_{P}(x_{1}) > V_{P}(q_{P})$, contradicting incentive compatibility of $q$. Thus, $q_{I}\big([x_{2},X]\big)>0$. Assume, towards a contradiction, that $x'_{2} < x_{2}$, so that $(f-\mu_{p}q_{p})([x_{1},x_{2}]) >0$. Consider allocation $q'$ with $q'_{P}=q_{P}$ and 
\begin{equation}
q'_{I} ~~~~~=~~~ q_{I} \cdot \mathbbm{1}\big\{1 \in [0,x_{1}] \big\} ~+~ \epsilon \cdot q_{I}\big((x_{2},X]\big) \cdot \big(f(x)-\mu_{P}q_{P}(x)\big) \cdot \mathbbm{1}\big\{ x \in (x_{1},x'_{2}) \big\} ~+~
\notag
\end{equation}
\begin{equation}
~+~ \Big(1-\epsilon \cdot (f-\mu_{P}q_{P})\big([x_{1},x'_{2}]\big) \Big ) \cdot q_{I}(x) \cdot \mathbbm{1}\big\{x \in [x_{2},X] \big\}.
\notag
\end{equation}
For small enough $\epsilon>0$, allocation $q'$ is feasible, incentive compatible, and a provides strict welfare improvement with respect to $q$, in contradiction. The claim follows. \hfill $\blacksquare$

\noindent \textbf{Proof of Lemma 4.} Let $q$ be an optimal allocation. Recall that an allocation $q$ exhibits disposal for type $k$ if there are sets $A,B$ such that $A \triangleleft B$, $(f - \mu_{P}q_{P}-\mu_{I}q_{I})(A)>0$, and $q_{k}(B)>0$. Although Lemma 3 shows that $P$-agents are served in one continuous block $[x_{1},x_{2}]$, this still leaves room for disposal involving $P$-agents, if there is some unused suppply on $[0,x_{1}]$. Thus, we formulate the next claim for both types:

\vspace{3mm}

\noindent \textbf{Claim A1.} If $IC_{ik}$ does not bind for $q$, then $q$ does not exhibit disposal for $k$-agents.

\noindent \textbf{Proof.} If $q$ exhibits disposal for $k$-agents, then for sufficiently small $\epsilon>0$, the allocation $q'$ with $q'_{i}=q_{i}$, and 
\begin{equation}
    q'_{k}(x) ~~~~~~~~~~~~=~~~~~~~ \Big ( q_{k}(x) +\epsilon \cdot q_{k}(B) \cdot \big(f(x)-\mu_{P}q_{P}(x)-\mu_{I}q_{I}(x)\big) \Big) \cdot \mathbbm{1}\big\{x \in A \big\} ~+~
    \notag
\end{equation}
\begin{equation}    
    ~~~~~~~~~~~+~ \Big(1-\epsilon \cdot(f-\mu_{P}q_{P}-\mu_{I}q_{I})(A)  \Big) \cdot q_{k}(x) \cdot \mathbbm{1} \big\{ x \in B \big\} ~+~ q_{k}(x) \cdot \mathbbm{1} \big\{x \not \in (A \cup B) \big\}
    \notag
\end{equation}
is feasible, incentive compatible, and provides a strict welfare improvement, in contradiction. \hfill $\square$

Assume $IC_{PI}$ does not bind. Then $q_{I}(\diamond)=0$, and $x_{3} = \sup ( \textrm{supp}(q)) \in [0,X]$. If $x_{3}\leq x_{2}$, then by Lemma 3, $x_{3} \leq x_{1}$, where $q_{P} = f ~\vert~ [x_{1},x_{2}]$. In this case, $V_{P}(q_{I})>u_{P}(x_{1})>V_{P}(q_{P})$, contradicting incentive compatibility. If $(f-\mu_{P}q_{P}-\mu_{I}q_{I})\big([0,x_{3}]\big)>0$, then for small enough $\delta >0$, we have $(f-\mu_{P}q_{P}-\mu_{I}q_{I})\big([0,x_{3}-\delta]\big)>0$. Notice also that $q_{I}\big((x_{3}-\delta,x_{3}]\big)>0$; otherwise, $x_{3} = \sup ( \textrm{supp}(q))\leq x_{3}-\delta < x_{3}$, in contradiction. Thus, the allocation $q$ exhibits disposal for type $I$ with $A=[0,x_{3}-\delta]$ and $B = (x_{3}-\delta,x_{3}]$, leading to a contradiction. 
Therefore, an optimal allocation in this case takes the form $q_{P} = f ~\vert~ [x_{1},x_{2}]$, $q_{I} = f ~\vert~ [0,x_{1}] \cup [x_{2},x_{3}]$, where $F(x_{3})=\mu_{P}+\mu_{I}$, consistent with the lemma's statement.

Assume $IC_{PI}$ binds. Then, $IC_{IP}$ does not bind by Corollary 2. By Lemma 3, $q_{P} = f ~\vert~ [x_{1},x_{2}]$. By our arguments above, $q$ does not exhibit disposal for $P$-agents. Thus, $q_{I}([0,x_{1}]) = \mu_{I}^{-1}F(x_{1})$. Denote $x_{3} = \sup \big(\textrm{supp}(q_{I}) \cap [0,X]\big)$. 

If $x_{3}=x_{1}$, then $q = (1-\beta) \cdot f ~\vert~ [0,x_{1}] + \beta \cdot \delta_{\diamond}$, which satisfies the lemma's statement. 

Otherwise, $x_{3}>x_{2}$. Towards a contradiction, assume that $(f-\mu_{I}q_{I})\big([x_{2},x_{3}]\big) >0$. Then, for small enough $\delta>0$, we have $x_{2}< x_{3}-\delta < x_{3}$, where $(f-\mu_{I}q_{I})(A) >0$ and $q_{I}(B) >0$ with $A = [x_{2},x_{3}-\delta]$. and $B = (x_{3}-\delta,x_{3}]$. By definition, $A \triangleleft B$.
Let $\gamma(x) \in (0,1)$ for $x \in B$ be such that 
\begin{equation*}
V_{P}\Big(\gamma(x) \cdot (f-q_{I}\mu_{I})~\vert~A + (1-\gamma(x)) \cdot \delta_{\diamond}\Big) = V_{P}(\delta_{x}).    
\end{equation*}
By Lemma 0, 
\begin{equation*}
V_{I}\Big(\gamma(x) \cdot (f-q_{I}\mu_{I})~\vert~A + (1-\gamma(x)) \cdot \delta_{\diamond}\Big) = V_{I}(\delta_{x}).    
\end{equation*}
Integrating these equations with respect to $q_{I} ~\vert~B $ yields
\begin{equation}
V_{P}\Big(\overline{\gamma} \cdot (f-q_{I}\mu_{I})~\vert~A + (1-\overline{\gamma}) \cdot \delta_{\diamond}\Big) ~=~ V_{P}(q_{I} ~\vert~B) ~~~,~~~
V_{I}\Big(\overline{\gamma} \cdot (f-q_{I}\mu_{I})~\vert~A + (1-\overline{\gamma}) \cdot \delta_{\diamond}\Big) ~>~ V_{I}(q_{I}~\vert~B)
    \notag
\end{equation}
for some $\overline{\gamma} \in (0,1)$. Consider allocation $q'$ given by $q'_{P}=q_{P}$ and 
\begin{equation}
    q'_{I} ~~~~=~~~~ q_{I} ~+~ \epsilon \cdot \big( \overline{\gamma} \cdot(f-q_{I}\mu_{I})~\vert~A + (1-\overline{\gamma}) \cdot \delta_{\diamond}\big) ~-~ \epsilon \cdot q_{I} ~\vert~ B.
    \notag
\end{equation}
For $\epsilon>0$ small enough, $q'$ is feasible, incentive compatible, and provides strict welfare improvement, in contradiction. Therefore, $q_{I} ~\vert~ [x_{2},x_{3}] = f ~\vert~ [x_{2},x_{3}]$, and $q_{I} = (1-\beta) \cdot f ~\vert~ [0,x_{1}]\cup [x_{2},x_{3}] + \beta \delta_{\diamond}$ and $q_{P} = f ~\vert~ [x_{1},x_{2}]$, satisfying the lemma's statement. \hfill $\blacksquare$

\vspace{3mm}

\noindent \textbf{Proof of Proposition 3.}
The optimal allocation exists by Proposition 0. Let $q$ be an optimal allocation. From the proof of Lemma 4, $q_{P} = f ~\vert~ [x_{1},x_{2}]$ and $q_{I} = (1-\beta) \cdot f ~\vert~ [0,x_{1}] \cup [x_{2},x_{3}] + \beta \cdot \delta_{\diamond}$. It remains to show that $q$ is a unique optimal allocation. Assume, towards a contradiction, that there is another optimal allocation $q'_{P} = f ~\vert~ [x'_{1},x'_{2}]$, $q'_{I} = (1-\beta') \cdot f ~\vert~ [0,x'_{1}] \cup [x'_{2},x'_{3}] + \beta' \cdot \delta_{\diamond}$. Consider the allocation $q'' = 0.5 \cdot q + 0.5 q'$. By linearity of the welfare function and the constraints, $q''$ is an optimal allocation as well.  Hence, $q''_{P} = f ~\vert~ [x''_{1},x''_{2}]$, $q''_{I} = (1-\beta'') \cdot f ~\vert~ [0,x''_{1}] \cup [x''_{2},x''_{3}] + \beta'' \cdot \delta_{\diamond}$, which is impossible unless $x_{i}=x'_{i}=x''_{i}$ for $i = 1,2,3$ and $\beta = \beta' = \beta''$, in which case $q''=q'=q$. To conclude the proof, feasibility requires $F(x_{2})-F(x_{1}) = \mu_{P}$, and $F(x_{3})-F(x_{2})+F(x_{1}) = (1-\beta) \mu_{I}$. Incentive compatibility for $I$-agents implies that $\beta<1$.  \hfill $\blacksquare$

\vspace{3mm}

\noindent \textbf{Proof of Proposition 4 and Corollary 4.} First, we show that if an allocation $q$ is feasible and does not exhibit disposal or an inverted spread, then $q \in \mathcal{A}$, where $\mathcal{A}$ is the set of allocations that are constructed recursively as described in Section 5.1. Absence of disposal means that $\sum_{i}\mu_{i}q_{i}(x) = f(x) \cdot \mathbbm{1}\big\{x \in [0,\overline{X}] \big\}$ and $q_{i}\big([0,\overline{X}]\big)=1$ for all $i$. 

\vspace{2mm}

\noindent \textbf{Claim B1.} Suppose allocation $q$ does not exhibit an inverted spread or disposal, and let $j<k$. Then $\nu \Big( \textrm{conv}\big(\textrm{supp}(q_{j})\big) \cap \textrm{supp}(q_{k}) \Big) = 0$.

\noindent \textbf{Proof.} Since $q$ does not exhibit disposal, $\diamond \not \in \textrm{supp}(q_{j})$ and $\textrm{conv}\big(\textrm{supp}(q_{j})\big)$ is well-defined. Let $x_{j} \equiv \inf(\textrm{supp}(q_{j}))$, and $x^{j} \equiv \sup(\textrm{supp}(q_{j}))$. Then $D \equiv \textrm{conv}(\textrm{supp}(q_{j})) \cap \textrm{supp}(q_{k})$  $=$ $[x_{j},x^{j}] \cap \textrm{supp}(q_{k})$. Towards a contradiction, assume that $\nu(D)>0$. Since $q_{k}(x) \leq \mu_{k}^{-1}\cdot f(x)$ is bounded above on $[x_{j},x^{j}]$, there exists $\epsilon>0$ such that $x_{j}+\epsilon < x^{j}-\epsilon$ and $q_{k}\big((x_{j}+\epsilon,x^{j}-\epsilon)\big)>0$. The definition of $x_{j},x^{j}$ implies that $q_{j}\big([x_{j},x_{j}+\epsilon]\big)>0$ and $q_{j}\big([x^{j}-\epsilon,x^{j}]\big)>0$. Therefore, $[x_{j},x_{j}+\epsilon] \triangleleft (x_{j}+\epsilon,x^{j}-\epsilon) \triangleleft [x^{j}-\epsilon,x^{j}]$ constitutes an inverted spread, in contradiction. \hfill $\square$

Let $x_{i} = \textrm{inf}(\textrm{supp}$ $(q_{i}))$ and  $x^{i} = \textrm{sup}(\textrm{supp}$ $(q_{i}))$ for $i=1,...,N$. Define recursively $f^{0}=f$ and $f^{k}=f^{k-1}-\mu_{k}q_{k}=f - \sum_{i=1}^{k}\mu_{i}q_{i}$ for $k=1,...,N-1$. It is easy to see that Claim B1 and  $\sum_{i}\mu_{i}q_{i}(x)=f(x)$ for $x \in [0,\overline{X}]$ imply that $q_{k}=f^{k-1} ~\vert~[x_{k},x^{k}]$ and $f^{k-1}\big([x_{k},x^{k}]\big)=\mu_{k}$. Thus, $q \in \mathcal{A}$.

\vspace{0.1 in}

Assume now that $q \in \mathcal{A}$. We have that $q_{1} = f ~\vert~ [x_{1},x^{1}]$ and $q_{k} = f^{k-1} ~\vert~ [x_{k},x^{k}]$, where $f^{k-1} = f - \sum_{i=1}^{k-1}\mu_{i}q_{i}$ and $f^{k-1}\big([x_{k},x^{k}]\big) = \mu_{k}$. This implies that $\big(f - \sum_{i=1}^{k} \mu_{i}q_{i}\big)\big([x_{k},x^{k}]\big)=0$. We now show that $q$ is a first-best allocation for some strictly positive welfare weights $(\alpha_{1},...,\alpha_{N})$. Define $w_{k} = \textrm{min}\{x_{k},...,x_{N} \}$ and $w^{k} = \textrm{max} \{x^{k},...,x^{N} \}$. 

Consider the following procedure that defines weights $\alpha_{1},...,\alpha_{N}$, auxiliary real numbers $b_{1},...,b_{N}$, and functions $v_{1},...,v_{N}:~[0,X] \rightarrow R$ recursively for $k=N,N-1,...,2,1$. 

\noindent For $k=N$, define $\alpha_{N}=1$, $b_{N}=0$, and $v_{N}(x)=u_{N}(x)$. 

\noindent For $k<N$, suppose we constructed our desired objects up to $k+1$. Consider the following three cases that we soon show are exhaustive:

\vspace{0.1 in}

\noindent \textbf{Case 1}: If $w_{k+1} < x_{k} < x^{k} < w^{k+1}$, define $\alpha_{k} = \dfrac{v_{k+1}(x_{k})-v_{k+1}(x^{k})}{u_{k}(x_{k})-u_{k}(x^{k})}$ and $b_{k} = v_{k+1}(x_{k})-\alpha_{k}u_{k}(x_{k})$. Then, $\alpha_{k}u_{k}(x_{k})+b_{k} = v_{k+1}(x_{k})$ and $\alpha_{k}u_{k}(x^{k})+b_{k} = v_{k+1}(x^{k})$.

\vspace{0.1 in}

\noindent \textbf{Case 2}: If $x_{k} < x^{k} \leq w_{k+1} < w^{k+1}$, define $\alpha_{k} ~=~ \dfrac{v_{k+1}(0)-v_{k+1}(x^{k})}{u_{k}(0)-u_{k}(x^{k})}$ and $b_{k} = v_{k+1}(0)-\alpha_{k}u_{k}(0)$. Then, $\alpha_{k}u_{k}(0)+b_{k} ~=~ v_{k+1}(0)$ and $\alpha_{k}u_{k}(x^{k})+b_{k} ~=~ v_{k+1}(x^{k})$.

\vspace{0.1 in}

\noindent \textbf{Case 3}: If $w_{k+1} < w^{k+1} \leq x_{k} < x^{k}$, define $\alpha_{k} ~=~ \dfrac{v_{k+1}(x_{k})-v_{k+1}(\overline{X})}{u_{k}(x_{k})-u_{k}(\overline{X})}$ and $b_{k} = v_{k+1}(x_{k})-\alpha_{k}u_{k}(x_{k})$. Then, $\alpha_{k}u_{k}(x_{k})+b_{k} ~=~ v_{k+1}(x_{k})$ and $\alpha_{k}u_{k}(\overline{X})+b_{k} ~=~ v_{k+1}(\overline{X})$. 

\vspace{0.1 in}

\noindent In all three cases, define
\begin{equation*}
v_{k}(x) \equiv \max \{v_{k+1}(x),\alpha_{k}u_{k}(x)+b_{k} \} = \textrm{max} \{ \alpha_{N}u_{N}(x) + b_{N}~,~ \alpha_{N-1}u_{N-1}(x) + b_{N-1}~,~...~,~\alpha_{k}u_{k}(x)+b_{k} \}.
\end{equation*}
Since $u_{k}(\cdot)$ is strictly decreasing, $\alpha_{k}$ is well-defined. The function $v_{k}(x)$ is also strictly decreasing, hence $\alpha_{k}>0$. Thus, $\alpha_{1},...,\alpha_{N}$ are conceivable welfare weights.

\vspace{3mm}

\noindent \textbf{Claim B2.} The three cases considered above are exhaustive.

\noindent \textbf{Proof.} Towards a contradiction, assume that $w^{k+1} \in (x_{k},x^{k}]$. Then there is some $j>k$ such that $\sup(\textrm{supp}(q_{j}))=x^{j}=w^{k+1}$, hence $q_{j}\big([x_{k},x^{k}]\big) = q_{j}\big([x_{k},x^{j})\big)>0$, contradicting $\big(f - \sum_{i=1}^{k} \mu_{i}q_{i}\big)\big([x_{k},x^{k}]\big)=0$ for $q \in \mathcal{A}$. Thus, $w^{k+1} \not \in (x_{k},x^{k}]$. A symmetric argument shows that $w_{k+1} \not \in [x_{k},x^{k})$. Since $w_{k}<w^{k}$, the statement of the Claim follows. \hfill $\square$

\noindent \textbf{Claim B3.} For all $k=N,N-1,...,2,1$, for all $j=k,k+1,...,N$: 
\begin{equation} 
 \Big \{~ x \in [w_{k},w^{k}] ~\Big \vert~ \alpha_{j}u_{j}(x) +b_{j} = v_{k}(x)  ~\Big \} ~~~=~~~
 [x_{j},x^{j}] ~\Big \backslash~ \left( \overset{j-1}{\underset{i=k}{\bigcup}}(x_{i},x^{i}) \right)
 \notag
\end{equation}
and $\alpha_{j}u_{j}(x) +b_{j} < v_{k}(x)$ for all $x \in [w_{k},w^{k}] \big \backslash \left( [x_{j},x^{j}] ~\Big \backslash~ \left( \overset{j-1}{\underset{i=k}{\bigcup}}(x_{i},x^{i}) \right) \right)$.

\noindent \textbf{Proof.} We use induction on $k=N,N-1,...,2,1$. 

For $k=N$, the claim is true since $v_{N}(x)=u_{N}(x)=\alpha_{N}u_{N}(x)+b_{N}$. 

Assume the claim holds up to $k+1$. Consider $\eta_{k}(x) ~=~ \alpha_{k} u_{k}(x) + b_{k} - v_{k+1}(x)$.
Let $x' < x''$ and $\lambda \in (0,1)$. Denote by $g_{kj}(x) = \alpha_{k}u_{k}(x)+b_{k}-\alpha_{j}u_{j}(x) - b_{j}$ for $j>k$. Type-$k$ agents are more risk averse than type-$j$ agents for any $j>k$. Since strict quasi-concavity is invariant with respect to positive affine transformations, Lemma 1 implies that $g_{jk}(\lambda x'+(1-\lambda)x'') > \textrm{min}\{g_{jk}(x'),g_{jk}(x'') \}$ and
\begin{equation}
    \eta_{k}(\lambda x' +(1-\lambda)x'') ~=~ 
\alpha_{k}u_{k}(\lambda x' +(1-\lambda)x'')+b_{k} ~- \underset{j=k+1,...,N}{\textrm{max}}\left\{\alpha_{j}u_{j}(\lambda x'+(1-\lambda)x'')+b_{j} \right\} ~=~
    \notag
\end{equation}
\begin{equation}
    ~=~ \underset{j=k+1,...,N}{\textrm{min}}\left(g_{kj}(\lambda x' +(1-\lambda)x'')\right) ~>~
    \underset{j=k+1,...,N}{\textrm{min}}\left(\textrm{min}\left\{g_{kj}(x'),g_{kj}(x'')\right\}\right) ~=~
    \notag
\end{equation}
\begin{equation}
    ~=~ \textrm{min}\left\{\underset{j=k+1,...,N}{\textrm{min}}\left(g_{kj}(x')\right)~,~\underset{j=k+1,...,N}{\textrm{min}}\left(g_{kj}(x'')\right)\right\} ~=~
    \textrm{min}\left\{ \eta_{k}(x'), \eta_{k}(x'') \right\}
    \notag
\end{equation}
We conclude that the function $\eta_{k}(x)$ is strictly quasi-concave. 

Consider case 1 above, corresponding to $w_{k+1} < x_{k} < x^{k} < w^{k+1}$. Then $w_{k}=w_{k+1}$ and $w^{k}=w^{k+1}$. We deal with the case of $j=k$ first. By construction, $\eta_{k}(x_{k})=\eta_{k}(x^{k})=0$. Hence, $\eta_{k}(\lambda x_{k}+(1-\lambda) x^{k})>0$ for $\lambda \in (0,1)$. We conclude that $v_{k}(x)=\alpha_{k}u_{k}(x)+b_{k}$ for $x \in [x_{k},x^{k}]$, and $v_{k}(x)>v_{k+1}(x)\geq \alpha_{j}u_{j}(x)+b_{j}$ for $j=k+1,...,N$ and $x \in (x_{k},x^{k})$. 

Assume, towards a contradiction, that $\eta_{k}(x) \geq 0$ for some $x \not \in [x_{k},x^{k}]$. If $x < x_{k}$ then, by strict quasi-concavity of $\eta_{k}$, we have that $\eta_{k}(x_{k})>0$, contradiction; similarly if $x > x^{k}$, then $\eta_{k}(x^{k})>0$, contradiction. The claim is true for $j=k$.  

Consider now $j>k$. From continuity of $v_{k}(\cdot)$ and our arguments above, $v_{k}(x)=v_{k+1}(x)$ for $x \not \in (x_{k},x^{k})$. Then,
\begin{equation} 
 \Big \{~ x \in [w_{k},w^{k}] ~\Big \vert~ \alpha_{j}u_{j}(x) +b_{j} = v_{k}(x)  ~\Big \} ~=~
 \Big \{~ x \in [w_{k+1},w^{k+1}] ~\Big \vert~ \alpha_{j}u_{j}(x) +b_{j} = v_{k}(x)  ~\Big \} ~=~
 \notag
\end{equation}
\begin{equation} 
  ~=~\Big \{~ x \in [w_{k+1},w^{k+1}]  ~\Big \vert~ \alpha_{j}u_{j}(x) +b_{j} = v_{k+1}(x)  ~\Big \} \Big \backslash (x_{k},x^{k}) ~=~ [x_{j},x^{j}] ~\Big \backslash~ \left(\left( \overset{j-1}{\underset{i=k+1}{\bigcup}}(x_{i},x^{i})  \right) \cup (x_{k},x^{k}) \right),
 \notag
\end{equation}
where the last equality follows from our induction hypothesis. The claim then follows.

Consider case 2 above, so that $x_{k}<x^{k}\leq w_{k+1}<w^{k+1}$. Then $w_{k}=x_{k}$ and $w^{k}=w^{k+1}$. By construction, $\eta_{k}(0)=\eta_{k}(x^{k})=0$. Repeating similar arguments to those used for case 1, we conclude that $v_{k}(x)=\alpha_{k}u_{k}(x)+b_{k} > v_{k+1}(x)$ for all $x \in (0,x^{k}) \supseteq (x_{k},x^{k})$ and $v_{k}(x) = v_{k+1}(x) > \alpha_{k}u_{k}(x)+b_{k}$ for all $x \in (x^{k},w^{k}] = [w_{k},w^{k}] \backslash [x_{k},x^{k}]$. Therefore, we can follow the same arguments used for case 1 to show the claim for $j=k,k+1,...,N$.

Finally, consider case 3, where $w_{k+1} < w^{k+1} \leq x_{k} < x^{k}$. Then $w_{k}=w_{k+1}$ and $w^{k}=x^{k}$. By construction, $\eta_{k}(x_{k})=\eta_{k}(\overline{X})=0$. Repeating the same arguments again, we conclude that $v_{k}(x)=\alpha_{k}u_{k}(x)+b_{k} > v_{k+1}(x)$ for all $x \in (x_{k},\overline{X}) \supseteq (x_{k},x^{k})$ and $v_{k}(x) = v_{k+1}(x) > \alpha_{k}u_{k}(x)+b_{k}$ for all $x \in [w_{k},x_{k}) = [w_{k},w^{k}] \backslash [x_{k},x^{k}]$. Therefore, we can repeat the arguments pertaining to case 1 above and the claim follows.\hfill $\square$

By Clam B3 for $k=1$, and the fact that $q \in \mathcal{A}$, it follows that for all $x \in [0,\overline{X}]$,
\begin{equation}
\begin{array}{lllllllll}
    a_{j}u_{j}(x) + b_{j} & = & v_{1}(x)  & \Longleftrightarrow & x \in  [x_{j},x^{j}] \Big \backslash \left( \overset{j-1}{\underset{i=1}{\bigcup}}(x_{i},x^{i}) \right) & \Longleftrightarrow & x \in \textrm{supp}(q_{j}), \\
    a_{j}u_{j}(x) + b_{j} & < & v_{1}(x)  & \Longleftrightarrow & x \not \in  [x_{j},x^{j}] \Big \backslash \left( \overset{j-1}{\underset{i=1}{\bigcup}}(x_{i},x^{i}) \right) & \Longleftrightarrow & x \not \in \textrm{supp}(q_{j}),
\end{array}
    \notag
\end{equation}
where, for any $q \in \mathcal{A}$, we use $(x_{j},x^{j}) \cap \textrm{supp}(q_{m}) = \varnothing$ for $m>j$  since $q_{m} = \big(f - \sum_{i<m}\mu_{i}q_{i}\big)$  $~\vert~ [x_{m},x^{m}]$. 

Consider welfare weights $\alpha_{1},...,\alpha_{N}$ and any allocation $q'$ that does not exhibit disposal.\footnote{As already noted, an allocation that exhibits disposal cannot be first-best---there is an obvious way to improve the welfare of such an allocation by assigning the unused supply of a better quality to an agent who exhibits disposal.} Then, 
 \begin{equation}
     W(q') ~+~\sum_{j=1}^{N}\mu_{j}b_{j}   ~=~
     \overset{\overline{X}}{\underset{0}{\int}} \sum_{j=1}^{N} \mu_{j}(\alpha_{j}u_{j}(x)+b_{j} )q'_{j}(x)dx ~\leq~ 
     \overset{\overline{X}}{\underset{0}{\int}} \sum_{j=1}^{N} \mu_{j}v_1(x)q'_{j}(x)dx ~\leq~  \overset{\overline{X}}{\underset{0}{\int}} v_{1}(x)f(x)dx ~=~ 
     \notag
 \end{equation}
 \begin{equation}
     ~=~ \sum_{j=1}^{N}\underset{\textrm{supp}(q_{j})}{\int} v_{1}(x)f(x)dx  ~=~ 
     \sum_{j=1}^{N}\underset{\textrm{supp}(q_{j})}{\int} (\alpha_{j}u_{j}(x)+b_{j})\mu_{j}q_{j}(x)dx ~=~  W(q) ~+~\sum_{j=1}^{N}\mu_{j}b_{j}
     \notag
 \end{equation}
We conclude that $q$ is a first-best allocation for strictly positive welfare weights $\alpha_{1},...,\alpha_{N}$.

The proof that any first-best allocation is Pareto efficient is standard and omitted. Suppose now that $q$ is a Pareto efficient allocation. As already argued, $q$  does not exhibit disposal. Suppose $q$ exhibits an inverted spread between types $j,k$ with $j<k$. We can fix the allocation of all other types $i \neq j,k$, and use an allocation $q'$ following the construction in the proof of Lemma 2 (with $P=j$ and $I=k$) to get a strict improvement for $k$-agents without altering other agents' payoffs. It follows that any Pareto efficient allocation does not exhibit disposal or an inverted spread. This completes the proof of Proposition 4.

To complete the proof of Corollary 4, two claims remain. First, we need to show that the number of different blocks is no more than $2N-1$. This follows from the fact that the boundaries of the blocks defining the allocation are $x_{N},...,x_{1},x^{1},...,x^{N}$, with some of these points possibly coinciding. Second, we need to show that the allocation given to $k$-agents consists of no more than $k$ disjoint blocks. This follows from $\textrm{supp}(q_{k}) = [x_{k},x^{k}] \Big \backslash \left( \overset{k-1}{\underset{i=1}{\bigcup}}(x_{i},x^{i}) \right)$. \hfill $\blacksquare$ 

\vspace{3mm}

\noindent \textbf{Proof of Proposition 5.}

\noindent By Proposition 0, a solution to the mechanism designer's problem exists. We show uniqueness after proving the first part of the proposition. 

\vspace{3mm}

\noindent \textbf{Proof of Part 1.} 
Suppose $q=(q_{1},...,q_{N})$ is a solution to the mechanism designer's problem. Consider two different agent types, $j$ and $k$. For $\epsilon >0$, consider the set 
\begin{equation}
A_{jk}^{\epsilon} ~\equiv~ \left\{~x \in [0,X]
~\Big\vert~ \text{min}\big\{q_{j}(x),q_{k}(x) \big\} >\epsilon ~\right\}.
\notag
\end{equation}
Since $q^{j}$ and $q^{k}$ are measurable, the set $A_{jk}^{\epsilon}$ is measurable as well. Let $\nu(\cdot)$ be a $\mathcal{L}$ebesque measure. Towards a contradiction, assume that  $\nu(A_{jk}^{\epsilon}) >0$. Since $\nu$ is non-atomic, we can partition $A_{jk}^{\epsilon}$ into $N+1$ disjoint subsets $\{ A_{i}\}_{i=1}^{N+1}$ such that $\nu(A_{i})>0$ for all $i=1,...,N+1$. Consider any arbitrary family of disjoint measurable sets $\{B_{i}\}_{i=1}^{N+1}$, with $B_{i} \subseteq A_{i}$ for all $i=1,...,N+1$. Denote by $B = \underset{i}{\bigcup} B_{i}$. Let $\omega ~=~ (\omega_{1},...,\omega_{N+1}) ~~\in~~ \left[0, 1 \right]^{N+1}$ and define the following allocation:
\begin{equation}
\begin{array}{llll}
\widetilde{q}_{j}(\omega)(x)  & = & q_{j}(x) \cdot \mathbbm{1}\{ t \not \in B \} ~+~ \left[q_{j}(x) - \epsilon + \epsilon \cdot 
\dfrac{\mu_{j}+\mu_{k}}{\mu_{j}} \cdot \displaystyle \sum_{i=1}^{N+1}(1 - \omega_{i}) \cdot \mathbbm{1}\{x \in B_{i}\}
\right] \cdot\mathbbm{1}\{x \in B\},\\
\widetilde{q}_{k}(\omega)(x) & = & q_{k}(x) \cdot \mathbbm{1}\{ t \not \in B \} ~+~ \left[q_{k}(x) - \epsilon + \epsilon \cdot 
\dfrac{\mu_{j}+\mu_{k}}{\mu_{k}} \cdot \displaystyle \sum_{i=1}^{N+1} \omega_{i} \cdot \mathbbm{1}\{x \in B_{i}\}
\right] \cdot\mathbbm{1}\{x \in B\},\\
\widetilde{q}_{l}(\omega)(x) & = & q_{l}(x) ~~~~~~~~~\text{for}~l \neq j,k.
\end{array}
\notag
\end{equation}
\noindent Define also 
\begin{equation} \label{appendix_h-definition}
\begin{array}{llllll}
h^{0}(\omega)  & \equiv &  \displaystyle \sum_{i=1}^{N+1}\omega_{i} \cdot \nu(B_{i}) ~~-~ \dfrac{\mu_{k}}{\mu_{j}+\mu_{k}} \cdot \nu(B)
 & = & \epsilon^{-1} \cdot \dfrac{\mu_{k}}{\mu_{j}+\mu_{k}} \cdot \big(\widetilde{q}_{k}([0,X] - q_{k}([0,X])\big),   \\
h^{l}(\omega) & \equiv &  \displaystyle \sum_{i=1}^{N+1}\left(\omega_{i}~-~\dfrac{\mu_{k}}{\mu_{j}+\mu_{k}}\right)\underset{B_{i}}{\displaystyle \int}u_{l}(x)dx
& = & \epsilon^{-1} \cdot \dfrac{\mu_{k}}{\mu_{j}+\mu_{k}} \cdot \big(V_{l}(\widetilde{q}_{k}(\omega)) - V_{l}(q_{k}) \big), 
\end{array}
\end{equation}
for all $l \in \{1,...,N \} $.

\vspace{2mm}

\noindent \textbf{Claim C1.} If $h^{0}(\omega)=0$, then the allocation $\widetilde{q}(\omega)$ is feasible.

\noindent \textbf{Proof.} By construction, $q_{j}(x),q_{k}(x)>\epsilon$ for $x \in B$. Thus, $\widetilde{q}_{i}(x) \geq 0$ for all $x \in [0,X]$ and $i \in \{1,...,n \}$. Additionally, $\mu_{j}\widetilde{q}_{j}+\mu_{k}\widetilde{q}_{k}=\mu_{j}q_{j}+\mu_{k}q_{k}$, implying that $f(x)-\sum_{i}\mu_{i}\widetilde{q}_{i}(x) = f(x)-\sum_{i}\mu_{i}q_{i}(x) \geq 0$ from $q$'s feasibility. It also implies that $\widetilde{q}_{j}([0,X]) = q_{j}([0,X])$ since $\widetilde{q}_{k}([0,X]) = q_{k}([0,X])$. Thus, $\widetilde{q}_{i}([0,X]) \leq 1$ for all $i \in \{1,...,N \}$, again using $q$'s feasibility. \hfill $\square$ 
\\
The functions $h^{l}(\omega)$ are linear in $\omega$, and their gradients are given by
\begin{equation*}
\begin{array}{lllllllll}
\triangledown_{\omega}h^{0} & = & \big(\nu(B_{1}) &,& ... &,&  \nu(B_{N+1})\big),  & \\   
 \triangledown_{w}h^{l}  & = & \left( \underset{B_{1}}{\displaystyle \int}u_{l}(x)dx \right. &,& ... &,& \left. \underset{B_{N+1}}{\displaystyle \int}u_{l}(x)dx \right), &~~~l=1,...,N.   
\end{array}
\end{equation*}

Let $\zeta^{l}_{m} = \left( \dfrac{\partial h^{l}}{\partial \omega_{1}}~,~\right.$ $...~,~$ $\left. \dfrac{\partial h^{l}}{\partial \omega_{m}}  \right) $ be a truncated gradient of $h^{l}$. \\

\noindent \textbf{Claim C2.} There exist $\{B_{i}\}_{i=1}^{N+1}$ as above such that the vectors $\zeta_{m}^{0},...,\zeta_{m}^{m-1}$ are linearly independent for $m \in \{1,...,N+1\}$.

\noindent \textbf{Proof.} We prove the claim by induction on $m=1,...,N+1$. 

For $m=1$, take $B_{1}=A_{1}$. Then, $\zeta_{1}^{0}=\nu(B_{1}) >0$.

Suppose the statement is true for $m$. Fix $B_{1},...,B_{m}$. Since $\zeta_{m}^{0},...,\zeta_{m}^{m-1}$ are linearly independent, these vectors constitute a basis in $\mathbb{R}^{m}$. Therefore, 
\begin{equation*}
\left( \dfrac{\partial h^{m}}{\partial \omega_{1}}~,\right....~, \left.~\dfrac{\partial h^{m}}{\partial \omega_{m}}  \right) = \overset{m-1}{\underset{i=0}{\sum}}\lambda_{i} \zeta_{m}^{i},    
\end{equation*}
where $\lambda_{i}$, $i=0,...,m-1$, are determined uniquely. 

Towards a contradiction, assume that for all $B_{m+1} \subseteq A_{m+1}$ such that $\nu(B_{m+1})>0$, the vectors $\zeta_{m+1}^{0},...,\zeta_{m+1}^{m}$ are not linearly independent. Then, 
\begin{equation*}
\zeta^{m}_{m+1} \equiv \left( \dfrac{\partial h^{m}}{\partial \omega_{1}}~,\right....~, \left.~\dfrac{\partial h^{m}}{\partial \omega_{m+1}}  \right) = \overset{m-1}{\underset{i=0}{\sum}}\lambda_{i} \zeta_{m+1}^{i},    
\end{equation*}
where $\lambda_{i}$, $i=0,...,m-1$, do not depend on $B_{m+1}$.  Hence, 
\begin{equation}
\underset{B_{m+1}}{\int} u_{m}(x)dx ~=~  \dfrac{\partial h^{m}}{\partial \omega_{m+1}} ~=~ \sum_{i=0}^{m-1} \lambda_{i} \cdot \dfrac{\partial h^{i}}{\partial \omega_{m+1}}  ~=~
\overset{m-1}{\underset{i=1}{\sum}}\lambda_{i} \underset{B_{m+1}}{\int} u_{i}(x)dx+\lambda_{0}\nu(B_{m+1})
\label{u_m}
\end{equation}
\noindent For any $\delta >0$, denote by 
\begin{equation*}
D^{+}_{\delta} \equiv \Big\{ x \in A_{m+1} ~\Big \vert~   u_{m}(x)> \overset{m-1}{\underset{i=1}{\sum}}\lambda_{i}  u_{i}(x) +\lambda_{0}+\delta \Big\},~ D^{-}_{\delta} \equiv \Big\{ x \in A_{m+1} ~\Big \vert~ u_{m}(x) < \overset{m-1}{\underset{i=1}{\sum}}\lambda_{i} u_{i}(x) + \lambda_{0} -\delta \Big\}.    
\end{equation*}
Then, 
\begin{equation*}
\underset{D_{\delta}^{+}}{\int} u_{m}(x)dx > \overset{m-1}{\underset{i=1}{\sum}}\lambda_{i} \underset{D_{\delta}^{+}}{\int} u_{i}(x)dx + \lambda_{0}\nu(D_{\delta}^{+}) + \delta \nu(D_{\delta}^{+}),~~~\textrm{and}   
\end{equation*}
\begin{equation*}
\underset{D_{\delta}^{-}}{\int} u_{m}(x)dx < \overset{m-1}{\underset{i=1}{\sum}}\lambda_{i} \underset{D_{\delta}^{-}}{\int} u_{i}(x)dx + \lambda_{0}\nu(D_{\delta}^{-}) - \delta \nu(D_{\delta}^{-}).    
\end{equation*}

If $\nu(D_{\delta}^{+})>0$ or $\nu(D_{\delta}^{-})>0$, we can pick $B_{m+1}=D_{\delta}^{+}$ or $B_{m+1}=D_{\delta}^{-}$, respectively, to achieve a contradiction through equation (\ref{u_m}). It follows that $\nu(D_{\delta}^{+})=\nu(D_{\delta}^{-})=0$.

It follows that
\begin{equation}
    \nu\left(\left\{  x \in A_{m+1} ~\Big \vert ~ u_{m}(x) = \overset{m-1}{\underset{i=1}{\sum}}\lambda_{i}  u_{i}(x)+\lambda_{0}
    \right\}\right) ~=~ \nu \left(A_{m+1} \Big \backslash \left( \overset{\infty}{\underset{r=1}{\bigcup}}\left(D_{1/r}^{+} \cup D_{1/r}^{-} \right) \right)\right) ~=~  \nu(A_{m+1}) >0,
    \notag
\end{equation}
in contradiction to the independence of different agent types' utility functions. \hfill $\square$ 

\vspace{2mm}

\noindent \textbf{Claim C3.} There exists a unit vector $e \in R^{N+1}$ such that $(\triangledown_{\omega}h^{k} \cdot e)$$>0$ and $(\triangledown_{\omega}h^{i} \cdot e) =0$ for $i \in \{0,1,...,N \} \backslash \{k \}$.\\
\noindent \textbf{Proof.} If $m=N+1$, then $\zeta_{m}^{l} = \triangledown_{\omega} h^{l}$. Thus, by Claim C2, the vector $\triangledown_{\omega} h^{k}$ is linearly independent of $\{ \triangledown_{\omega} h^{i} \}_{i \in \{0,1,...,N \} \backslash \{k \} }$. The claim follows. \hfill $\square$\\

\noindent Denote by $\omega^{*}=\frac {\mu_{k}} {\mu_{j}+\mu_{k}} \cdot(1,...,1) \in [0,1]^{N+1}$ and define $\omega = \omega^{*} + c \cdot e \in [0,1]^{N+1}$ for $c>0$ small enough, say, $c = (1/2) \cdot \min \big\{\frac {\mu_{k}} {\mu_{j}+\mu_{k}},\frac {\mu_{j}} {\mu_{j}+\mu_{k}} \big\}$.

\vspace{2mm}

\noindent \textbf{Claim C4.} Let $\widetilde{q}=\widetilde{q}(\omega)$ be an allocation defined as above. Then, (1) $V_{k}(\widetilde{q}_{k}) > V_{k}(q_{k})$; (2) $V_{k}(\widetilde{q}_{j}) < V_{k}(q_{j})$; (3) $V_{k}(\widetilde{q}_{l}) = V_{k}(q_{l})$ for all $l \in \{1,...,n \} \backslash \{j,k \}$; and (4)   $V_{r}(\widetilde{q}_{l})=V_{r}(q_{l})$ for all $r \in \{1,...,n\} \backslash \{k\}$, $l \in \{1,...,n\}$. \\
\textbf{Proof.} Recall that $\widetilde{q}(\omega^{*})=q$ and $h^{l}(\omega^{*})=0$, for all $l \in \{0,1,...,N\}$. Thus,
\begin{equation*}
V_{k}(\widetilde{q}_{k}) - V_{k}(q_{k}) = \epsilon \cdot \frac{\mu_{j}+\mu_{k}}{\mu_{k}} \cdot c \cdot (\triangledown_{\omega}h^{k} \cdot e) >0,    
\end{equation*}
so statement (1) holds. Next, $\mu_{j} \widetilde{q}_{j}+\mu_{k} \widetilde{q}_{k} = \mu_{j}q_{j}+\mu_{k}q_{k}$. Therefore,
\begin{equation*}
V_{k}(\widetilde{q}_{j}) = V_{k}\Big( \frac {\mu_{j}q_{j}+\mu_{k}q_{k}-\mu_{k}\widetilde{q}_{k}}{\mu_{j}} \Big) = V_{k}(q_{j})+\frac{\mu_{k}} {\mu_{j}} \cdot \Big( V_{k}(q_{k}) - V_{k}(\widetilde{q}_{k})\Big) < V_{k}(q_{j}),    
\end{equation*}
proving statement (2). Statement (3) follows from $\widetilde{q}_{l}(\omega) = q_{l}$ for $l \neq j,k$. We now show statement (4). If $l \in \{1,...,n \} \backslash \{j,k \}$, then statement (4) follows from $\widetilde{q}_{l}(\omega) = q_{l}$. From Claim C3, $(\triangledown_{\omega}h^{r} \cdot e) =0$ for all $r \neq k$. Therefore, $h^{r}(\omega)=  h^{r}(\omega^{*}) + \epsilon \cdot \frac{\mu_{j}+\mu_{k}}{\mu_{k}} \cdot c \cdot (\triangledown_{\omega}h^{r} \cdot e) =0$ for all $r \neq k$. Thus, if $l=k$, then $V_{r}(\widetilde{q}_{k}) = V_{r}(q_{k})+ h^{r}(\omega) = V_{r}(q_{k})$. Finally, for $l=j$, 
\begin{equation*}
V_{r}(\widetilde{q}_{j}) = V_{r}\big(\frac {\mu_{j}q_{j}+\mu_{k}q_{k}-\mu_{k}\widetilde{q}_{k}}{\mu_{j}} \big) = V_{r}(q_{j})+\frac{\mu_{k}} {\mu_{j}} \cdot\big(V_{r}(q_{k})-V_{r}(\widetilde{q}_{k}) \big)=V_{r}(q_{j}).    
\end{equation*}
\hfill $\square$

\noindent \textbf{Claim C5.} Allocation $\widetilde{q}(\omega)$, constructed above, is feasible and incentive compatible.\\
\textbf{Proof.} Feasibility follows from Claim C1 since, from Claim C3, $h^{0}(\omega) = h^{0}(\omega^{*})+\epsilon \cdot \frac{\mu_{j}+\mu_{k}}{\mu_{k}} \cdot c \cdot (\triangledown_{\omega}h^{0} \cdot e) = 0$. Incentive compatibility follows from Claim C4 and incentive compatibility of $q$. 
\hfill $\square$

By Claim C4, 
\begin{equation*}
W(\widetilde{q}) = \sum_{i=1}^{n}\mu_{i}V_{i}(\widetilde{q}_{i}) = \sum_{i \neq k}\mu_{i}V_{i}(q_{i}) + \mu_{k}V_{k}(\widetilde{q}_{k}) > \sum_{i=1}^{n}\mu_{i}V_{i}(q_{i}) =W(q).    
\end{equation*}
Using Claim C5, the allocation $\widetilde{q}$ is a viable improvement and $q$ cannot be optimal. Therefore, $\nu(A^{\epsilon}_{jk})=0$ for all $\epsilon>0$ for all $j,k \in \{1,...,n \}$ such that $j \neq k$. It follows that, at the optimal allocation, supply can be used by at most one type of agent almost everywhere.\\

We now show that, whenever a supply of some quality good is used non-trivially, its supply is exhausted. The proof mirrors our construction above. Namely, for any $\epsilon>0$, we now consider 
\begin{equation*}
A_{0k}^{\epsilon} ~\equiv~ \left\{~x \in [0,X] ~\Big\vert~ \text{min}\big\{q_{k}(x),f(x)-\sum_{i}\mu_{i}q_{i}(x) \big\} >\epsilon ~\right\}.     
\end{equation*}
Towards a contradiction, assume that $\nu(A_{0k}^{\epsilon})>0$. Introduce a partition $\{A_{i} \}_{i=1}^{N+1}$ of $A_{0k}^{\epsilon}$, and then corresponding subsets $\{B_{i} \}_{i=1}^{N+1}$ and $B$ as in our analysis of $A^{\epsilon}_{jk}$. Let $\omega \in [0,1]^{N+1}$, and let $q'(\omega)$ be given by
\begin{equation}
q'_{k}(\omega)(x) ~=~ q_{k}(x) \cdot \mathbbm{1}\{ t \not \in B \} ~+~ \left[q_{k}(x) - \epsilon + \epsilon \cdot 
\frac{1+\mu_{k}}{\mu_{k}} \overset{N+1}{\underset{i=1}{\sum}} \omega_{i} \cdot \mathbbm{1}\{x \in B_{i}\}
\right] \cdot\mathbbm{1}\{x \in B\}, 
\notag
\end{equation}
with $q'_{l}=q_{l}$ for $l \neq k$. We then have 
\begin{equation}
   \epsilon^{-1} \cdot \frac{\mu_{k}}{1+\mu_{k}} \big(q'_{k}([0,X] - q_{k}([0,X])\big) ~=~ h^{0}(\omega) ~~~~~~,~~~~~~\epsilon^{-1} \cdot \frac{\mu_{k}}{1+\mu_{k}} \big(V_{l}(q'(\omega)_{k})-V_{l}(q_{k}) \big) ~=~ h^{l}(\omega)
    \notag
\end{equation}
for $l = 1,...,N$, where $h^{i}(\omega)$, $i=0,1,...,N$ are defined in eq. (\ref{appendix_h-definition}).\\

\noindent \textbf{Claim C1'.} If $h^{0}(\omega)=0$, the allocation $q'(\omega)$ is feasible.\\
\textbf{Proof.} Since $q_{k}(x)>\epsilon$ for $x \in B$, and $q'_{l} =q_{l}$ for $l \neq k$, then $q'_{i}(x) \geq 0$ for all $x \in [0,X]$ and $i \in \{1,...,n \}$. Next, since
$f(x)-\sum_{i}\mu_{i}q_{i}(x)>\epsilon$ for $x \in B$, then $f(x)-\sum_{i}\mu_{i}q'_{i}(x) \geq f(x)-\sum_{i}\mu_{i}q_{i}(x) -\epsilon \geq 0$ for $x \in B$, and $f(x)-\sum_{i}\mu_{i}q'_{i}(x) = f(x)-\sum_{i}\mu_{i}q_{i}(x) \geq 0$ for $x \not \in B$. Finally, $q'_{k}([0,X])=q_{k}([0,X]) +\epsilon \cdot h^{0}(\omega)=q_{k}([0,X])$, so that $q'_{i}([0,X])=q_{i}([0,X]) \leq 1$ for all $i$. \hfill $\square$ 
 
By Claim C3, there exists a unit vector $e \in R^{N+1}$ such that $(\triangledown_{\omega}h^{k} \cdot e)$ $>0$ and $(\triangledown_{\omega}h^{i} \cdot e) =0$ for $i \in \{0,1,...,N \} \backslash \{k \}$. Denote by $\omega^{*}=\frac {\mu_{k}} {1+\mu_{k}} \cdot(1,...,1) \in [0,1]^{N+1}$, and consider $\omega = \omega^{*} + c \cdot e\in [0,1]^{N+1}$, where $c>0$ is small enough, say, $c = (1/2) \cdot \min \big\{\frac{\mu_{k}}{1+\mu_{k}},\frac{1}{1+\mu_{k}} \big\}$.

\vspace{2mm}

\noindent \textbf{Claim C4'.} Let $q'=q'(\omega)$ be defined as above. Then, (1) $V_{k}(q'_{k}) > V_{k}(q_{k})$; (2) $V_{k}(q'_{l}) = V_{k}(q_{l})$ for all $l \in \{1,...,n \} \backslash \{k \}$; and (3)   $V_{r}(q'_{l})=V_{r}(q_{l})$ for all $r \in \{1,...,n\} \backslash \{k\}$, $l \in \{1,...,n\}$ \\
\textbf{Proof.} Recall that $q'(\omega^{*})=q$ and $h^{l}(\omega^{*})=0$ for all $l \in \{0,1,...,N\}$. Thus, 
\begin{equation*}
V_{k}(q'_{k}) - V_{k}(q_{k}) = \epsilon \cdot \frac{1+\mu_{k}}{\mu_{k}} \cdot c \cdot (\triangledown_{\omega}h^{k} \cdot e) >0,    
\end{equation*}
so statement (1) holds. For $l \neq k$, statements (2) and (3) follow from $q'_{l}(\omega) = q_{l}$. Finally, since  $(\triangledown_{\omega}h^{r} \cdot e) =0$ for all $r \neq k$, we have $V_{r}(q'_{k}) = V_{r}(q_{k})+ h^{r}(\omega) = V_{r}(q_{k})$. \hfill $\square$ 

\noindent \textbf{Claim C5'.} Allocation $q'(\omega)$, constructed above, is feasible and incentive compatible.\\
\textbf{Proof.} Feasibility follows from $h^{0}(\omega) = h^{0}(\omega^{*})+\epsilon \cdot \frac{1+\mu_{k}}{\mu_{k}} \cdot c \cdot (\triangledown_{\omega}h^{0} \cdot e) = 0$ and Claim C1'. Incentive compatibility follows from Claim C4'. \hfill $\square$ 

By Claim C4', 
\begin{equation*}
W(q') = \sum_{i=1}^{n}\mu_{i}V_{i}(q'_{i}) = \sum_{i \neq k}\mu_{i}V_{i}(q_{i}) + \mu_{k}V_{k}(q'_{k}) > \sum_{i=1}^{n}\mu_{i}V_{i}(q_{i}) =W(q). 
\end{equation*}
Using Claim C5', we conclude that allocation $q'$ is feasible, incentive compatible, and provides a strict welfare improvement over $q$, contradicting $q$'s optimality. Therefore,  $\nu(A^{\epsilon}_{0k})=0$ for all $\epsilon>0$ and all $k$. Hence, at the optimal allocation, for almost all $x$, either the entire supply is utilized or none of it is.

\vspace{3mm}

\noindent \textbf{Proof of uniqueness.} Assume, towards a contradiction, that there are two different optimal allocations $q$ and $q'$. Since neither $q$ nor $q'$ have mass points, they differ on a set $A \subseteq [0,X]$ of positive $\mathcal{L}$ebesque measure. Since the mechanism design problem entails a linear objective subject to linear constraints, the set of optimizers is convex. In particular, $0.5q+0.5q'$ is also an optimal allocation. However, $0.5q+0.5q'$ violates our conclusions above. Indeed, for any $x \in A$, we have $q(x) \neq q'(x)$. It follows that there is $k \in \{1,...,N\}$ such that $q_{k}(x) \neq q'_{k}(x)$. Since $0 \leq q_{k}(x) \leq \frac{f(x)}{\mu_{k}}$, and $0 \leq q'_{k}(x) \leq \frac{f(x)}{\mu_{k}}$, we have $0 < 0.5q_{k}(x)+0.5q'_{k}(x) < \frac{f(x)}{\mu_{k}}$. Because the number of types is finite, this happens for a positive measure of $x \in A$ for at least one agent type, contradicting what we already showed.

\vspace{3mm}

\noindent \textbf{Proof of Part 2.} Assume, towards a contradiction, that an optimal allocation $q$ exhibits a directed cycle in the graph of binding IC constraints. That is, there exist a sequence of types, $k_{1},...,k_{m}$, such that $V_{k_{1}}(q_{k_{1}})=V_{k_{1}}(q_{k_{2}})~,~V_{k_{2}}(q_{k_{2}})=V_{k_{2}}(q_{k_{3}})~,~...~,~V_{k_{m}}(q_{k_{m}})=V_{k_{m}}(q_{k_{1}})$. Consider the following allocation $q'$: 
\begin{equation}
q'_{k_{i}}~=~\left( 1-\epsilon/\mu_{k_{i}} \right) q_{k_{i}}+\left(\epsilon/\mu_{k_{i}}\right)q_{k_{i+1}}~~~~,
~~~~\textrm{where}~~~~\epsilon = \left(1/2 \right) \cdot \underset{j}{\text{min}}(\mu _{j})~>0,  
\notag
\end{equation}
\begin{equation}
q'_{j}~=~q_{j}~~~~\text{for}~j\neq k_{1},...,k_{m},  \notag
\end{equation}
where we define $k_{m+1}\equiv k_{1}$. The allocation $q'_{l}$ is a convex combination of $q_{j}$ for all $l$ because of our choice of $\epsilon$. It is straightforward to see that the allocation $q'$ is feasible. Next, 
\begin{equation}
V_{j}(q'_{j})~=~V_{j}(q_{j})~~~~\text{for}~j\neq k_{1},...,k_{m}  \notag
\end{equation}
and
\begin{equation}
V_{k_{i}}(q'_{k_{i}})~=~V_{k_{i}}\left(\left( 1-\epsilon/\mu_{k_{i}} \right) q_{k_{i}}+\left(\epsilon/\mu_{k_{i}}\right)q_{k_{i+1}}\right) =
\left( 1-\epsilon/\mu_{k_{i}} \right) V_{k_{i}}(q_{k_{i}})+\left(\epsilon/\mu_{k_{i}}\right)V_{k_{i}}(q_{k_{i+1}})=V_{k_{i}}(q_{k_{i}}). 
\notag
\end{equation}
Therefore, the payoffs of all agent types under the allocation $q'$ are the same as under the allocation $q$. The resulting welfare is then identical. 

Now, for an arbitrary type $l$ we have: 
\begin{equation}
V_{l}(q'_{l})~=~V_{l}(q_{l})~\geq ~V_{l}(q_{j})~=~V_{l}(q'_{j})~~~~\text{for}~j\neq k_{1},...,k_{m},  \notag
\end{equation}
\begin{equation}
V_{l}(q'_{l})~=~V_{l}(q_{l}) ~\geq~ \left( 1-\epsilon/\mu_{k_{i}} \right) V_{l}(q_{k_{i}})+\left(\epsilon/\mu_{k_{i}}\right)V_{l}(q_{k_{i+1}})~=~
V_{l}(q'_{k_{i}}). 
\notag
\end{equation}
We conclude that $q'$ is also incentive compatible and, therefore, optimal. However, since $q_{k_{i}} \neq q_{k_{i+1}}$ as a consequence of the proposition's first part proven earlier, $q' \neq q$. This contradicts the uniqueness of an optimal allocation, already shown. \hfill $\blacksquare$ 

\vspace{3mm}

\noindent \textbf{Proof of Corollary 5.} Proposition 2 proven above is a special case of Corollary 5 when $N=2$. Thus, we assume $N>2$. We identify an incentive-compatible allocation $q^{*}$ for which none of the $IC$ constraints is binding, and $q^{*}$ is a first-best solution for some welfare weights $\alpha^{*}$ (now, a vector). The statement of the proposition then follows directly from Berge's theorem---the first-best allocation is continuous with respect to the welfare weights and therefore so are the functions $V_{j}(q_{j})-V_{j}(q_{k})$ for all $j,k$. Therefore, for any vector of weights $\alpha$ in a small enough neighborhood of $\alpha^{*}$, we have $V_{j}(q_{j})-V_{j}(q_{k}) >0$ for all $j \neq k$ for the corresponding first-best allocation $q$.

We call any pair of types $i,i+1$ for $i=1,..,N-1$ \emph{adjacent}. We say that a feasible allocation $q$ exhibits an \emph{accordion structure for adjacent types $i,i+1$}, if $q_{i+1} =  f ~\vert~ [x_{i+1},x_{i}] \cup [x^{i},x^{i+1}]$, and $q_{i} = f ~\vert~ [x_{i},x_{i-1}] \cup [x^{i-1},x^{i}]$ for some $0 \leq x_{i+1} \leq x_{i} \leq x_{i-1} \leq x^{i-1} \leq x^{i} \leq x^{i+1}$. A feasible allocation $q$ exhibits an \emph{accordion structure} if it exhibits an accordion structure for any adjacent pairs of types. In this case, there are $0=x_{N}\leq x_{N-1} \leq...\leq x_{1} < x^{1} \leq ... \leq x^{N-1} \leq x^{N}=\overline{X}$ such that $q_{i} = f ~\vert~ [x_{i},x_{i-1}] \cup [x^{i-1},x^{i}]$ and $q_{1} = f ~\vert~ [x_{1},x^{1}]$, where  $x_{0}=x^{0}=\frac{x_{1}+x^{1}}{2}$.

\vspace{2mm}

\noindent \textbf{Claim F1.} Let $A \triangleleft B \triangleleft C$ for non-empty measurable subsets $A,B,C \subseteq [0,\overline{X}]$. Then, for each type of agent $i$, there is a unique number $\gamma =\gamma_{i}(A,B,C) \in (0,1)$ such that an $i$-type agent is indifferent between $\gamma \cdot f ~\vert~A + (1-\gamma) \cdot f ~\vert~C$ and $f ~\vert~B$. Moreover, if $j>i$, then $\gamma_{j}(A,B,C) < \gamma_{i}(A,B,C)$.\\
\textbf{Proof.} Existence and uniqueness follow from the continuity and strict monotonicity of $V_{i} \big(\gamma \cdot f ~\vert~A + (1-\gamma) \cdot f ~\vert~C  \big)$ with respect to $\gamma \in [0,1]$ and the fact that $V_{i}(f ~\vert~A) >V_{i}(f ~\vert~ B) > V_{i}(f ~\vert~ C) $. In fact, we can directly identify
\begin{equation*}
    \gamma_{i}(A,B,C)  ~=~ \int \gamma_{i}(A,\delta_{x},C) d(f~\vert B).
\end{equation*}
By Lemma 0, $\gamma_{j}(A,\delta_{x},C) < \gamma_{i}(A,\delta_{x},C)$ for all $x \in B$. It follows that $\gamma_{j}(A,B,C) < \gamma_{i}(A,B,C)$.
\hfill $\square$

\vspace{2mm}

\noindent \textbf{Claim F2.} Suppose $q$ has an accordion structure for adjacent types $i,i+1$. Then at least one of the constraints $IC_{i(i+1)}$ and $IC_{(i+1)i}$ does not bind.\\
\textbf{Proof.} Let $A=(x_{i+1},x_{i})$, $B=(x_{i},x_{i-1}) \cup (x^{i-1},x^{i})$, and $C=(x^{i},x^{i+1})$.  If $A=\varnothing$, then $V_{i}(q_{i}) > V_{i}(q_{i-1})$, and if $C = \varnothing$, then $V_{i+1}(q_{i+1})>V_{i+1}(q_{i})$. In these cases, the statement of the claim holds. 

Suppose $A,C \neq \varnothing$. From $q$'s feasibility, $B \neq \varnothing$. Thus, $A \triangleleft B \triangleleft C$. Define $\beta$ by 
\begin{equation*}
\beta \cdot f ~\vert~ [x_{i+1},x_{i}] + (1-\beta) \cdot f ~\vert~ [x^{i},x^{i+1}] = f ~\vert~ [x_{i+1},x_{i}] \cup [x^{i},x^{i+1}].   
\end{equation*}
Thus, $\beta = \dfrac{f\big([x_{i+1},x_{i}]\big)}{f\big([x_{i+1},x_{i}] \cup [x^{i},x^{i+1}]\big)} $. If $V_{i}(q_{i}) \leq V_{i}(q_{i+1})$, then $\beta  \leq \gamma_{i}(A,B,C) < \gamma_{i+1}(A,B,C)$, where we use Claim F1 and the fact that $f ~\vert~ A = f ~\vert~ [x_{i+1},x_{i}]$, $f ~\vert~ B = f ~\vert~ [x_{i},x_{i-1}] \cup [x^{i-1},x^{i}] $, and $f ~\vert~ C = f ~\vert~ [x^{i},x^{i+1}]$. Therefore, $V_{i+1}(q_{i}) < V_{i+1}(q_{i+1})$. Otherwise, $V_{i}(q_{i}) > V_{i}(q_{i+1})$. In both cases the claim's statement holds. \hfill $\square$

\noindent \textbf{Claim F3.} If $q$ has an accordion structure and all constraints $IC_{i(i+1)}$, $IC_{(i+1)i}$ for $i=1,...,N-1$ do not bind, then $q$ is incentive compatible, and no incentive constraint binds.\\
\textbf{Proof.} Define $\beta_{i}$ as above: 
\begin{equation*}
\beta_{i} \cdot f ~\vert~ [x_{i+1},x_{i}] + (1-\beta_{i}) \cdot f ~\vert~ [x^{i},x^{i+1}] = f ~\vert~ [x_{i+1},x_{i}] \cup [x^{i},x^{i+1}]. 
\end{equation*}
Let $A_{i} \equiv (x_{i+1},x_{i})$, $C_{i} \equiv (x^{i},x^{i+1})$, and $B_{i} \equiv (x_{i},x_{i-1}) \cup (x^{i-1},x^{i}) = A_{i-1} \cup C_{i-1}$. Since $IC_{i(i+1))}$ and $IC_{i(i+1)}$ do not bind, $\gamma_{i+1}(A_{i},B_{i},C_{i})  < \beta_{i} <\gamma_{i}(A_{i},B_{i},C_{i})$ for all $i =1,...,N-1$. Consider any types $j<k$. By Claim F1, $\gamma_{k}(A_{m},B_{m},C_{m}) < \beta_{m}$ for all $m<k$. Therefore, $V_{k}(q_{k}) > V_{k}(q_{k-1})>...>V_{k}(q_{j})$. Similarly, $\gamma_{j}(A_{m},B_{m},C_{m}) > \beta_{m}$ for all $m \geq j$. Thus, $V_{j}(q_{j}) > V_{j}(q_{j+1})>...>V_{j}(q_{k})$, as needed. \hfill $\square$

We use a recursive procedure to define a parametric family of allocations $\{ q^{x_{1}} \}_{x_{1} \in Y}$, where $Y = [0,F^{-1}(F(\overline{X})-\mu_{1})]$, that have an accordion structure. Let $k \in \{ 1,...,N \}$ denote the state of the procedure. The procedure starts at state $k=1$ and ends at state $k = N$. 

If $k=1$, define $x^{1}=F^{-1}(F(x_{1})+\mu_{1})$ 
and proceed to state $2$.

If $k \in \{2,...,N-1 \}$, assume that $x_{k-1},...,x_{1},x^{1},...,x^{k-1}$ have already been defined. There are then three cases.

First, if $F(x_{k}) < \mu_{k+1}$ and $V\big( f ~\vert~ [0,x_{k}] \cup [x^{k},x'] \big) \leq V_{k+1}(q_{k})$, where $x' = F^{-1}\big( \sum_{i=1}^{k}\mu_{i} \big)$. In this case, define $x_{k}=x_{k+1}=...=x_{N-1}=0$, $x^{j} = F^{-1}\big(\sum_{i=1}^{j}\mu_{i} \big)$ for $j=k,...,N-1$ and proceed to state $N$.

Second, $F(\overline{X})-F(x^{k-1}) \leq \mu_{k}$ and $V_{k-1} \big( f ~\vert~ [x'',x_{k-1}] \cup [x^{k-1},\overline{X}]  \big) \geq V_{k-1}(q_{k-1})$, where $x'' = F^{-1}\big(\sum_{i=k+1}^{N}\mu_{i} \big)$. In this case, define $x^{k} =x^{k+1}=...=x^{N-1}=\overline{X}$, $x_{j}=F^{-1} \big( \sum_{i=j+1}^{N}\mu_{i}  \big)$ for $j=k,k+1,...,N-1$ and proceed to state $N$.

Otherwise, let $y_{k}$ and $y^{k}$, with $0<y_{k}<y^{k}<\overline{X}$, be the unique quality levels satisfying 
\begin{equation*}
    V_{k-1}\big( f ~\vert~ [y_{k},x_{k-1}] \cup [x^{k-1},y^{k}] \big) ~=~ V_{k-1}(q_{k-1})   ~~~~~~,~~~~~
    f\big([y_{k},x_{k-1}] \cup [x^{k-1},y^{k}] \big) ~=~ \mu_{k}.
\end{equation*}
Similarly, let $z_{k}$ and $z^{k}$ be the unique quality levels satisfying 
\begin{equation*}
    V_{k}\big( f ~\vert~ [z_{k},x_{k-1}] \cup [x^{k-1},z^{k}] \big) ~=~ V_{k}(q_{k-1})   ~~~~~~,~~~~~
    f\big([z_{k},x_{k-1}] \cup [x^{k-1},z^{k}] \big) ~=~ \mu_{k}.
\end{equation*}
Define $x_{k} = \frac{y_{k}+z_{k}}{2}$, $x^{k} = F^{-1}(F(x^{k-1})+\mu_{k}-F(x_{k-1})+F(x_{k}))$, and proceed to state $k+1$.

If $k=N$, define $x_{N}=0$, $x^{N}=\overline{X}$ and end the procedure.

It follows directly that $x_{N},x_{N-1},$ $...,$ $x_{2},$ $x^{1},...,x^{N}$ and the corresponding accordion allocations are continuous in $x_{i} \in Y$. Let $A = \big\{ x_{1} \in Y ~\big\vert~ V_{N-1}(q^{x_{1}}_{N-1}) > V_{N-1}(q^{x_{1}}_{N}) \big\}$ and $B = \big\{ x_{1} \in Y ~\big\vert~ V_{N}(q^{x_{1}}_{N}) > V_{N}(q^{x_{1}}_{N-1}) \big\}$. If $x_{1} = 0$, then $q_{N-1} = f ~\vert~ [x^{N-2},x^{N-1}]$ and $q_{N}= f ~\vert~ [x^{N-1},\overline{X}]$. Therefore, $0 \in A \neq \varnothing$. Similarly, if $x_{1} = F^{-1}(F(\overline{X})-\mu_{1})$, then $q_{N-1} = f ~\vert~ [x_{N-1},x_{N-2}]$ and $q_{N} = f ~\vert~ [0,x_{N-1}]$. Therefore, $F^{-1}(F(\overline{X})-\mu_{1}) \in B \neq \varnothing$. Since $V_{i}(q^{x_{1}}_{j})$ for all $i,j$ are continuous in $x_{1}$, it follows that $A$ and $B$ are open sets. By Claim F2, $A \cup B = Y$. Since $Y$ is a connected set, there is $x^{*}_{1} \in A \cap B \neq \varnothing$. 
 
Finally, we show that $q^{x^{*}_{1}}$ is an allocation as  desired. If the procedure with parameter $x^{*}_{1}$ encounters the first case in its specification for some $k \in \{2,...,N-1 \}$, then $q_{N-1} = f ~\vert~ [x^{N-2},x^{N-1}]$ and $q_{N}= f ~\vert~ [x^{N-1},\overline{X}]$, contradicting $x^{*}_{1} \in B$. If the procedure encounters the second case in its specification for some $k \in \{2,...,N-1 \}$, then $q_{N-1} = f ~\vert~ [x_{N-1},x_{N-2}]$ and $q_{N} = f ~\vert~ [0,x_{N-1}]$, contradicting $x^{*}_{1} \in A$. Consider then an arbitrary state of the procedure, $k \in \{2,...,N-1 \}$. 
For arbitrary $w_{k}<x_{k}$, denote by $w^{k} = F^{-1}(F(x^{k-1})+\mu_{k}+F(x_{k-1})-F(w_{k}))$. Then $V_{j}\big( f ~\vert~ [w_{k},x_{k-1}] \cup [x^{k-1},w^{k}] \big)$ is strictly decreasing in $w_{k}$ for any $j$. By Claim F2, $y_{k} < z_{k}$. Therefore, $y_{k} < x_{k} < z_{k}$, and we conclude that both $IC_{k(k-1)}$ and $IC_{(k-1)k}$ are satisfied and not binding. By our choice of $x^{*}_{1}$, we also know that both $IC_{N(N-1)}$ and $IC_{(N-1)N}$ do not bind. By Claim F3, the allocation $q^{x^{*}_{1}}$ is incentive compatible. Since $q^{x^{*}_{1}}$ has an accordion structure, $q^{x^{*}_{1}} \in \mathcal{A}$. By Proposition 4 and Corollary 4, there are positive welfare weights $\alpha$ such that $q^{x^{*}_{1}}$ is a first-best solution, concluding the proof. \hfill $\blacksquare$

\vspace{3mm}

\noindent \textbf{Proof of Proposition 6.} For brevity, we use the shorthand of  ``equilibrium'' to represent ``fair competitive equilibrium.'' The proof that an equilibrium exists follows standard arguments and, for completeness, provided in the Online Appendix (Lemma A3). We now show the asserted resulting structure of any equilibrium allocation. 

Consider an $i$-agent's problem. Let $\eta_{i}$ be the Lagrange multiplier of the feasibility constraint $1-q_{i}\big([0,\overline{X}]\big) \geq 0$, and $\xi_{i}$ be the Lagrange multiplier of the budget constraint. The Lagrange function for agent $i$'s problem is
\begin{equation}
\mathcal{L}(q_{i}) ~=~ \int_{0}^{\overline{X}} \big[u_{i}(x) - \eta_{i} - \xi_{i} p(x)\big]dq_{i}(x) + \xi_{i} \omega_{i} + \eta_{i},
\notag
\end{equation}
where the optimization of $\mathcal{L}$ is over all distributions $q_{i}$ on Borel subsets of $[0,\overline{X}]$. Denote
\begin{equation} \label{p_i_competitive}
p^{i}(x) ~~=~~ \frac{1}{\xi_{i}} \cdot u_{i}(x) - \frac{\eta_{i}}{\xi_{i}}. 
\notag
\end{equation}
If $(p,q)$ is an equilibrium, then $\xi_{i}>0$. Indeed, if $\xi_{i}=0$, then either $u_{i}(0)-\eta_{i}>0$, in which case the problem has no solution, or $u_{i}(x) -\eta_{i}<0$ for all $x \in (0,\overline{X}]$, in which case $q_{i}\big((0,\overline{X}] \big)=0$. In both cases market-clearing fails. Hence, $p^{i}(x)$ is well-defined. It follows that $1/\xi_{i} > 0$ and $\eta_{i}/\xi_{i} \geq 0$. We can write $i$-agents' Lagrange function as follows:
\begin{equation}
\mathcal{L}(q_{i}) ~=~ \xi_{i} \cdot \int_{0}^{\overline{X}} \left[p^{i}(x)- p(x)\right]dq_{i}(x) + \xi_{i} \omega_{i} + \eta_{i}.
\notag
\end{equation}

\vspace{2mm}

\noindent \textbf{Claim D1.} Suppose $(p,q)$ is an equilibrium, then: (1) $p(x) \geq p^{i}(x)$ for all $x \in [0,\overline{X}]$; (2) $B_{i} \equiv \big\{ x \in [0,\overline{X}]  ~\big\vert~ p(x)=p_{i}(x)  \big\} \neq \varnothing$; and (3) $q_{i}(B_{i})=1$.\\

\textbf{Proof.} Assume that $p(x)<p^{i}(x)$ for some $x \in [0,\overline{X}]$, then the problem of the Lagrangian maximization has no solutions, since substituting $q_{i} = l \cdot \delta_{x}$ with $l=1,2,..$ yields an unbounded sequence of values of the Lagrange function. Thus, statement (1) holds. Assume that $p(x)>p^{i}(x)$ for all $x \in [0,\overline{X}]$, then the solution is $q_{i}=0$, violating the market-clearing condition. Therefore, statement (2) holds. Finally, it is never optimal to choose $q_{i}(x)>0$ for $x$ such that $p(x)>p^{i}(x)$. Thus, $q_{i}\big( [0,\overline{X}] \backslash B_{i} \big) =0$. The market-clearing condition implies $q_{i}(B_{i})=1$. \hfill $\square$ 

Consider
\begin{equation}
g_{jk}(x) ~~\equiv~~ p^{j}(x) - p^{k}(x) ~~=~~ \frac{1}{\xi_{j}}  u_{j}(x) - \frac{1}{\xi_{k}} u_{k}(x) - \left(\frac{\eta_{j}}{\xi_{j}}-\frac{\eta_{k}}{\xi_{k}} \right)
\notag
\end{equation}
From the proof of Lemma 1, it follows that $g_{jk}(x)$ is strictly quasi-concave for $j<k$. Denote by $\underline{x}_{j}=\textrm{inf}\big( \textrm{supp}(q_{j}) \big)$ and $\overline{x}_{j}=\textrm{sup}\big(\textrm{supp}(q_{j}) \big)$.

\vspace{2mm}

\noindent \textbf{Claim D2.} If $k>j$, then $q_{k}\big([\underline{x}_{j},\overline{x}_{j}]\big)=0$\\
\textbf{Proof.} Let $x \in (\underline{x}_{j},\overline{x}_{j})$. There exist $x',x''$ such that $\underline{x}_{j}<x' <x <x''<\overline{x}_{j}$ and  $p^{j}(x')=p(x')$, $p^{j}(x'')=p(x'')$. Otherwise, Claim D1 would imply that $q_{j}\big((\underline{x}_{j},x) \big) = 0$ or $q_{j}\big((x,\overline{x}_{j}) \big) = 0$, contradicting the definition of $\underline{x}_{j}$ and $\overline{x}_{j}$. By Claim D1, we get $p^{k}(x') \leq p(x')=p^{j}(x')$ and $p^{k}(x'') \leq p(x'')=p^{j}(x'')$. Hence, $g_{jk}(x') \geq 0$ and $g_{jk}(x'') \geq 0$. Since $g_{jk}(x)$ is strictly quasi-concave, $g_{jk}(x)>0$. It follows that $p^{k}(x) < p^{j}(x) \leq p(x)$. Therefore, $q_{k}(x) = 0$ by Claim D1. \hfill $\square$

\vspace{2mm}

\noindent \textbf{Claim D3.} For any $k=1,...,N$, if $(q,p)$ is an equilibrium, then $\underline{x}_{k}< \underline{x}_{k-1}<...<\underline{x}_{1} < \overline{x}_{1} <...< \overline{x}_{k-1} < \overline{x}_{k}$, $q^{1} = f ~\big\vert~ [\underline{x}_{1},\overline{x}_{1}]$ and $q_{j} = f ~\big\vert~ [\underline{x}_{j},\underline{x}_{j-1}] \cup [\overline{x}_{j-1},\overline{x}_{j}]$ for $j \in \{2,...,k\}$. \\
\textbf{Proof.} We prove the Claim by induction on $k$. 

\noindent For $k=1$, Claim D2 and the market-clearing condition imply that $q^{1} = f ~\big\vert~ [\underline{x}_{1},\overline{x}_{1}]$. 

\noindent Suppose the statement holds for $k$. By Claim D2, $q_{k+1}\big([\underline{x}_{k},\overline{x}_{k}] \big)=0$. Therefore, there are three possible cases.  

\noindent First, suppose $\underline{x}_{k+1}<\overline{x}_{k+1} \leq \underline{x}_{k} < \overline{x}_{k}$. Any agent prefers lottery $q_{k+1}$ to lottery $q_{k}$---by the market-clearing condition, $q_{m}\big([\underline{x}_{m},\overline{x}_{m}] \big)=1$ for any $m$. Since all agents have the same endowment, $q_{k+1}$ is feasible for a type-$k$ agent, contradicting the optimality of $q_{k}$.

\noindent Second, suppose $\underline{x}_{k} < \overline{x}_{k} \leq \underline{x}_{k+1}<\overline{x}_{k+1}$. As in the first case, agent $k+1$ makes a suboptimal choice.

\noindent Therefore, it must be that $\underline{x}_{k+1} < \underline{x}_{k}  < \overline{x}_{k} < \overline{x}_{k+1}$. Claim D2 and the market-clearing condition implies that $q_{k+1} = f ~\big\vert~ [\underline{x}_{k+1},\underline{x}_{k}] \cup [\overline{x}_{k},\overline{x}_{k+1}]$. \hfill $\square$ 

Claim D3 for $k=N$ and the market-clearing condition imply the proposition. \hfill $\blacksquare$

\bibliographystyle{ecta}
\bibliography{library}

\end{document}